\documentclass{article}

 \usepackage{siunitx} 

\usepackage[dvips]{graphicx} 
\usepackage{latexsym}
\usepackage{amsmath}
\usepackage{amsthm}
\usepackage{mathrsfs}
\usepackage[numbers]{natbib}
\let \chapter \section
\usepackage[ruled, vlined, linesnumbered]{algorithm2e}
\usepackage{enumerate}
\usepackage{enumitem}
\usepackage{multirow}
\usepackage{color}
\usepackage{xfrac}
\usepackage{lmodern}

\usepackage{amssymb}
\usepackage{booktabs}
\usepackage{mathrsfs}
\usepackage{enumerate}
\usepackage{color}
\usepackage{enumitem}
\usepackage{pbox}
\usepackage{rotating}
\usepackage{url}
\usepackage{authblk}

\newcommand{\MaxEntBiSet}{MERCER\xspace}

\newcommand{\abs}[1]{\left|{#1}\right|}

\newcommand{\schema}{S}
\newcommand{\rows}{\mathit{rows}}
\newcommand{\argmax}{\operatornamewithlimits{argmax}}

\newtheorem{lemma}{Lemma}

\title{Interactive Discovery of Coordinated Relationship Chains \\
with Maximum Entropy Models}

\author{Hao Wu\textsuperscript{1},
Maoyuan Sun\textsuperscript{1},
Peng Mi\textsuperscript{1},
Nikolaj Tatti\textsuperscript{2}, \\
Chris North\textsuperscript{1},
Naren Ramakrishnan\textsuperscript{1}
}
\affil{\textsuperscript{1}Discovery Analytics Center, Virginia Tech, USA \\
\textsuperscript{2}HIIT, Department of Information and Computer Science, Aalto University, Finland}

\date{}

\begin{document}

\maketitle

\begin{abstract}
Modern visual analytic tools promote human-in-the-loop analysis but are limited
in their ability to direct the user toward interesting and promising directions
of study. This problem is especially acute when the analysis task is exploratory
in nature, e.g., the discovery of potentially coordinated relationships in
massive text datasets. Such tasks are very common in domains like intelligence
analysis and security forensics where the goal is to uncover surprising
coalitions bridging multiple types of relations. We introduce new maximum
entropy models to discover surprising chains of relationships leveraging count
data about entity occurrences in documents. These models are embedded in a
visual analytic system called \MaxEntBiSet that treats relationship bundles as
first class objects and directs the user toward promising lines of inquiry. We
demonstrate how user input can judiciously direct analysis toward valid
conclusions whereas a purely algorithmic approach could be led astray.
Experimental results on both synthetic and real datasets from the intelligence
community are presented.
\end{abstract}

\section{Introduction}
\label{sec:intro}

\begin{figure}[!t]
  \centering
  \includegraphics[width=1\textwidth]{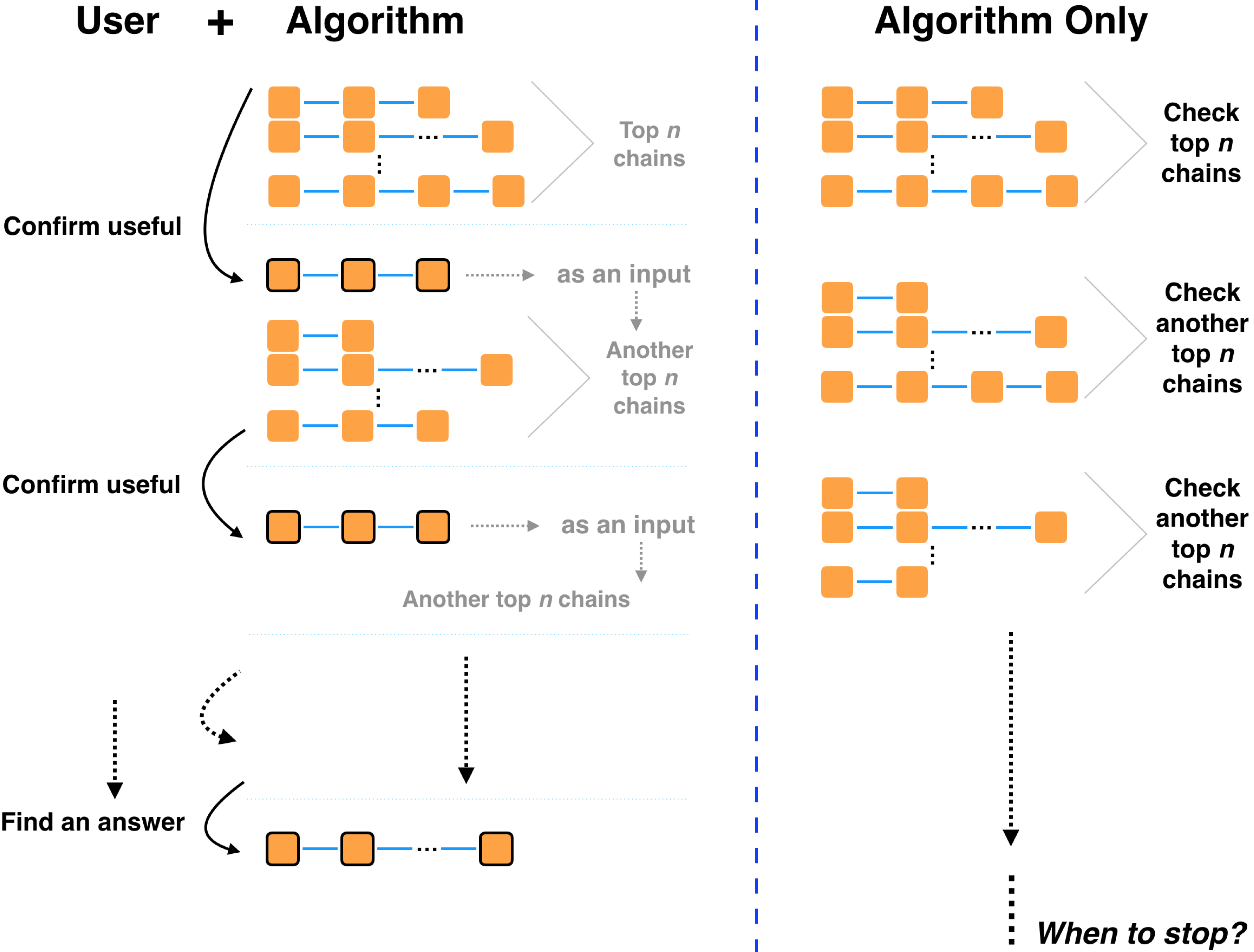}
  \caption{Illustration of \MaxEntBiSet. (left) Discovery of coordinated
relationship chains is aided by regular incorporation of user feedback.
(right) Unaided algorithmic discovery of relationship chains leads to
long lists of patterns that might not lead to the desired answer.}\label{fig:intro}
\end{figure}

Unstructured exploration of large text datasets is a crucial problem in many
application domains, e.g., intelligence analysis, biomedical discovery, analysis
of legal briefs and opinions. The state-of-the-art today involves two broad
classes of techniques. Visual analytic tools, e.g.,
Jigsaw~\cite{stasko2008jigsaw}, support the exploration of relationships
extracted from large text datasets. While they promote human-in-the-loop
analysis, identifying promising leads to explore is left to the creativity of
the user. At the other end of the spectrum, text analysis techniques such as
storytelling~\cite{Hossain:2012:SEN:2339530.2339742} provide interesting
artifacts (e.g., stories, summaries) for analysis but are limited in their
ability to incorporate user input to steer the discovery process.

Our goal here is to realize an amalgamation of algorithmic and human-driven
techniques to support the discovery of coordinated relationship chains. A
coordinated relationship (also called a bicluster) is one in which a group of
entities are related to another group of entities via a common relation. It is
thus a generalization of a relationship instance. A chain of such coordinated
relationships enables us to bundle groups of entities across various domains and
relate them through a succession of individual relationships.  The primary
artifact of interest are thus chains summarizing how entities in a document
collection are related. We introduce new maximum entropy (MaxEnt) models to
identify surprising chains of interest and rank them for inspection by the user.
In intelligence analysis, such chains can reveal how hitherto unconnected people
or places are related through a sequence of intermediaries. In biomedical
discovery, such chains can reveal how proteins involved in distinct pathways are
related through cross-talk via other proteins or signaling molecules. In legal
briefs, one can use chains to determine how rationale for court opinions vary
over the years and are buttressed by the precedence structure implicit in legal
history.

As shown in Fig.~\ref{fig:intro} (left), we envisage an interactive approach
wherein user feedback is woven at each stage and used to rank the most
interesting chains for further exploration. We will demonstrate through case
studies how such an approach gets users to their intended objectives faster than
a purely algorithmic approach (Fig.~\ref{fig:intro} (right)).  The work
presented here is implemented in a system -- Maximum Entropy Relational Chain
ExploRer (\MaxEntBiSet) that uses a variety of visual exploration strategies and
algorithmic means to foster user exploration.

Our key contributions are:
\begin{enumerate}
	\item \MaxEntBiSet is a marriage of two of our prior
		works~\cite{Wu:2014:UPD:2664051.2664089, sun2015biset} but supercedes
		the state-of-the-art in these papers in significant, orthogonal, ways. \MaxEntBiSet
		is a significant improvement over the work presented
		in~\cite{sun2015biset} because~\citet{sun2015biset} provides support for
		only manual exploration of coordinated relationships. \MaxEntBiSet is a
		significant improvement over the work presented
		in~\cite{Wu:2014:UPD:2664051.2664089}
		because~\citet{Wu:2014:UPD:2664051.2664089} only presents approaches to
		rank chains involving a binary maximum entropy model whereas
		\MaxEntBiSet introduces more general maximum entropy approaches for
		real-valued data.

\item We present two path strategies (full path and stepwise) to help analyze
	datasets. Using our proposed maximum entropy models, the full path strategy 
	discovers the most surprising bicluster chains from all possible chains
	involving an analyst-selected bicluster. The stepwise strategy evaluates
	biclusters neighboring a user-specified one, and prioritizes possible
	connected information with the current pieces under investigation. Both strategies
	directs analysts to reveal hidden plots involving surprising relational
	patterns.

\item We describe new visual encodings and summary as well as detailed views to
	support user-guided exploration of coordinated relationships in massive
	datasets. Besides basic color codings (e.g., \textsl{connection-oriented}
	highlighting presented in ~\cite{sun2015biset}), \MaxEntBiSet offers highlighting
	mechanisms aimed at pointing out surprising information.
Enhanced with the proposed maximum entropy models, this highlighting capability not only
	directs user's attention to important connected pieces of information, but
	also visually prioritizes them in a usable manner. 

\item We describe experimental results on both large, synthetic datasets (to
	illustrate efficiency and effectiveness of our algorithms) and small, real
	datasets (to illustrate how users can interactively explore a realistic text
	dataset. In particular, we show how \MaxEntBiSet enables the user to more
	quickly arrive at plots of interest than the traditional manual approach
	described in~\cite{sun2015biset}.
\end{enumerate}

\section{Preliminaries}
\label{sec:prelim}
\begin{figure}[!t]
	\centering
	\includegraphics[width=1.0\textwidth]{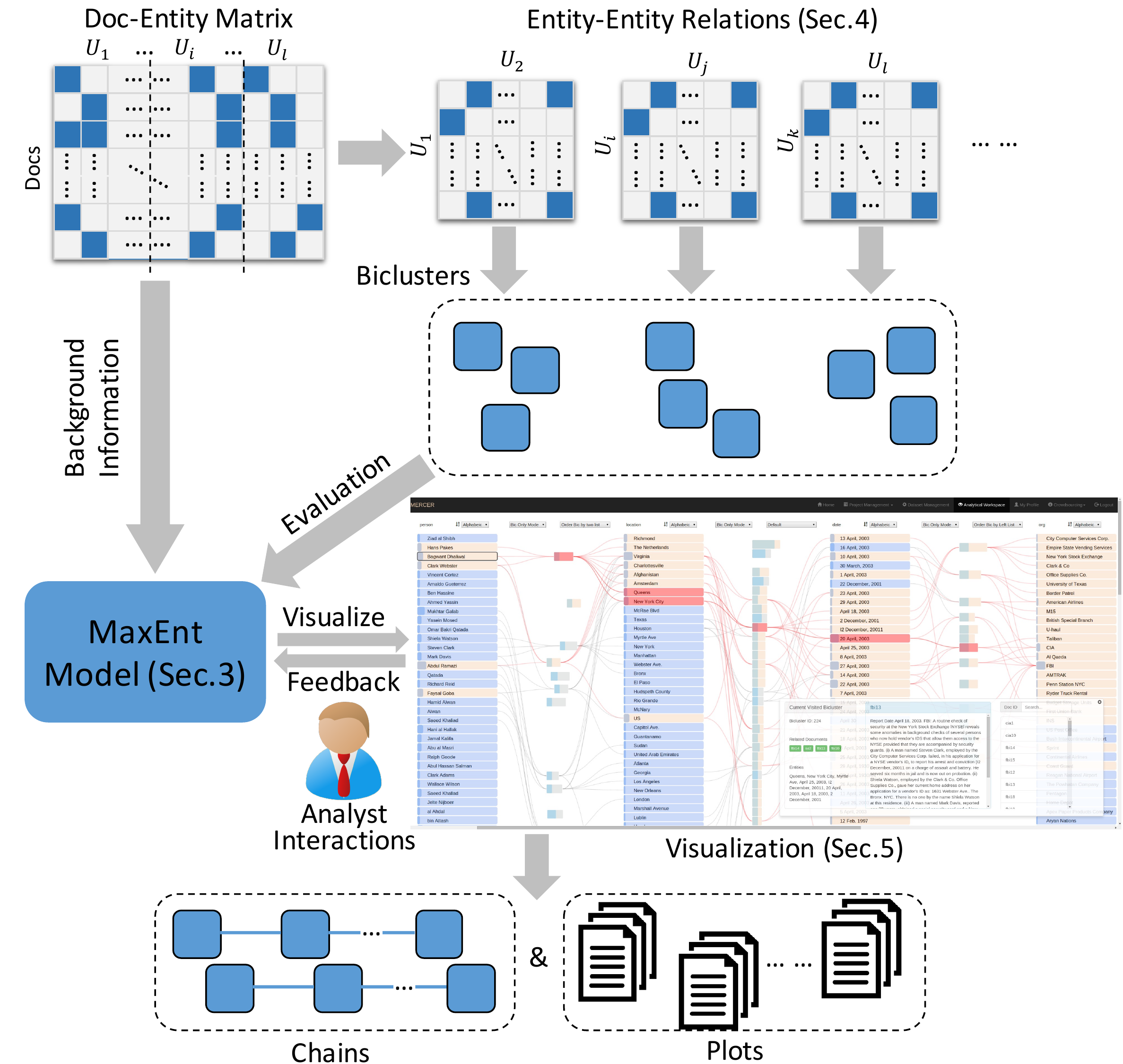}
	\caption{\MaxEntBiSet system workflow.}
	\label{fig:mercer_arch}
\end{figure}

Figure~\ref{fig:mercer_arch} illustrates the workflow in \MaxEntBiSet.
By taking the background information from the document-entity transactional
matrix, the \MaxEntBiSet system infers the maximum entropy model, which will be
described in detail in Section~\ref{sec:model}. From the document-entity matrix,
multiple entity-entity relations are extracted and surprisingness measure for
relational patterns is defined based on the MaxEnt model
(Section~\ref{sec:score}). By interacting with analysts, our visualization
interface displays the surprising relational patterns discovered from the
multiple entity-entity relations, and also provides analysts' feedback to the
MaxEnt model, which will in turn help to further discover additional surprising
patterns (Section~\ref{sec:biset}).

In this section, we introduce some preliminary concepts and notations that will
be useful to understand the \MaxEntBiSet system and the rest of this paper.

\paragraph{Multi-relational schema}
We assume that we are given $l$ domains or \emph{universes} which we will denote
throughout the paper by $U_i$, $i \in 1\ldots l$. An entity is a member of $U_i$
and an entity set $E_i$ is simply a subset of $U_i$.  We use $R = R(U_i, U_j)$
to denote a binary relation between some $U_i$ and $U_j$.  Given a set of
domains $\mathcal{U} = \{U_1, U_2,\ \ldots, U_l\}$ and a set of relations
$\mathcal{R} = \{R_1, R_2, \ \ldots, R_m\}$, a \emph{multi-relational schema}
$S(\mathcal{U}, \mathcal{R})$ is a connected bipartite graph whose vertex set is
given by $\mathcal{U} \cup \mathcal{R}$ and edge set is the collection of edges
each of which connects a relation $R_j$ in $\mathcal{R}$ and a domain $U_i$ in
$\mathcal{U}$ that the relation $R_j$ involves. In this paper, without loss of
generality, all vertices in $\mathcal{R}$ are assumed to have degree of two,
i.e., only binary relationships are considered. As is well known, ternary and
higher-order relations can be converted into sets of binary relationships. (No
such degree constraint exists for $\mathcal{U}$; a domain can participate in
many relationships.) 



\paragraph{Tiles}
A \emph{tile} $T$, a notion introduced by~\citet{Geerts04tilingdatabases}, is
essentially a rectangle in a data matrix. Formally, it is defined as a tuple $T
= (r(T), c(T))$ where $r(T)$ is a set of row identifiers (e.g., row IDs) and
$c(T)$ is a set of column identifiers (e.g., column IDs) over the data matrix. In
this most general form, it imposes no constraints on values of the matrix
elements identified by a tile. So, each element in a tile could be any valid
value in the data matrix. 
In the binary case, when all elements within a tile $T$ have the same value
(i.e., either all $1$s or all $0$s) we say it is an exact tile. Otherwise we
call it a noisy tile.

\paragraph{Biclusters}
As local patterns of interest over binary relations, we consider biclusters. A
\emph{bicluster}, denoted by $B = (E_i, E_j)$, on relation $R = R(U_i, U_j)$,
consists of two entity sets $E_i \subseteq U_i$ and $E_j \subseteq U_j$ such
that $E_i \times E_j \subseteq R$. As such a bicluster is a special case of an
exact tile, one in which all the elements are 1. Further, we say a bicluster $B
= (E_i, E_j)$ is {\it closed} if for every entity $e_i \in U_i \setminus E_i$,
there is some entity $e_j \in E_j$ such that $(e_i, e_j) \notin R$ and for every
entity $e_j \in U_j \setminus E_j$, there is some entity $e_i \in E_i$ such that
$(e_i, e_j) \notin R$.  In other words, $E_i$ is maximal (w.r.t. $E_j$) so that
we cannot add more elements to $E_i$ without violating the premise of a
bicluster. If a pair of entities $e_i \in U_i, e_j \in U_j$ belongs to a
bicluster $B$, we denote this fact by $(e_i, e_j) \in B$.


\paragraph{Redescriptions}
Assume that we are given two biclusters $B = (E_i, E_j)$ and $C = (F_j, F_k)$,
where $E_i \subseteq U_i$, $E_j, F_j \subseteq U_j$, and $F_k \subseteq U_k$.
Note that $E_j$ and $F_j$ lie in the same domain. Assume that we are given a
threshold $0 \leq \varphi \leq 1$. We say that $B$ and $C$ are \emph{approximate
redescriptors} of each other, which we denote by $B \sim_{\varphi, j} C$ if the
Jaccard coefficient $\abs{E_j \cap F_j} / \abs{E_j \cup F_j} \geq \varphi$. The
threshold $\varphi$ is a user parameter, consequently we often drop $\varphi$
from the notation and write $B \sim_j C$.  The index $j$ indicates the common
domain over which we should take the Jaccard coefficient. If this domain is
clear from the context we often drop $j$ from the notation. 
If $B \sim_{1, j} C$, then we must have $E_j = F_j$ in which case we say that
$B$ is an \emph{exact redescription} of $C$.

This definition is a generalization of the definition given
by~\citet{Zaki:2005:RSU:1081870.1081912}, who define redescriptions for
itemsets over their mutual domain, transactions, such that the set $E_j$
consists of transactions containing itemset $E_i$ and the set $F_j$ consists
of transactions containing itemset $F_k$.

\paragraph{Bicluster Chains}
A \emph{bicluster chain} $C$ consists of an ordered set of biclusters $\{B_{1},
B_{2}, \ldots, B_{k}\}$ and an ordered bag of domain indices $\{j_1, j_2,
\ldots, j_{k - 1}\}$ such that for each pair of adjacent biclusters we have
$B_{i} \sim_{j_i} B_{i + 1}$. Note that this implicitly requires that two
adjacent biclusters share a common domain.
If a bicluster $B_{R_i}$ is a part of the bicluster chain $C$, we will represent
this by $B_{R_i} \in C$ in this paper.

\paragraph{Surprisingness}
In the knowledge discovery tasks studied here, the primary goal is to extract
novel, interesting, or unusual knowledge. That is, we aim to find results that
are highly informative with regard to what we already know---we are not so much
interested in what we already do know, or what we can trivially induce from such
knowledge.
To this end, we suppose a probability distribution $p$ that represents the
user's current beliefs about the data. When mining the data (e.g., for a
bicluster or chain), we can use $p$ to determine the likelihood of a result
under our current beliefs: if it is high, this indicates that we most likely
already know about it, and thus, reporting it to the user would provide little
new information. In contrast, if the likelihood of a result is very low, the
result is very surprising, thus potentially conveying new information. In
Section~\ref{sec:model}, we will discuss how to infer such a probability
distribution for both binary and real-valued data matrices.

\paragraph{Problem Statement}
Given a multi-relational dataset, a bicluster chain across multiple relations
describes a progression of entity coalitions. We are particularly interested in
chains that are surprising w.r.t.\ what we already know, as these could help to
uncover the plots hidden in the multi-relational dataset.

More formally, given a multi-relational dataset schema $\schema(\mathcal{U},
\mathcal{R})$, where $\mathcal{U} = \{U_1, U_2, \ldots, U_l\}$ and
$\mathcal{R} = \{R_1, R_2, \ldots, R_m\}$, we are interested in iteratively finding
non-redundant bicluster chains that are most surprising with regard to each
other and w.r.t.\ the background knowledge with the assistance of visual
analysis techniques.


\section{Tile-based Maximum Entropy Model}
\label{sec:model}
Our problem statement is based on a notion of a multi-relational schema. In practice,
such multi-relational datasets are inferred from a transactional dataset (e.g., entities
discovered from a document collection, and then subsequently related by co-occurrence).
More specifically, we assume that our schema was generated from a transactional
data matrix $D$ (see Fig.~\ref{fig:mercer_arch}). This data matrix can be viewed
as a matrix of size $N$-by-$M$. We will introduce the method of obtaining a
schema from $D$ in Section~\ref{sec:score}. In this approach the columns of $D$
correspond to the entities of the schema. Hence, we will refer to the columns of
$D$ as entities.

\subsection{Maximum Entropy Model for Binary Data}
\label{sec:binary_maxent}
In this section, we will define the maximum entropy (MaxEnt) model for binary
data matrices using tiles as background knowledge---recall that a tile is a more
general notion than a bicluster. We will first introduce notation that will be
useful to understand the model derivation in the context of binary data. Then,
we will recall MaxEnt theory for modeling binary data given tiles as background
information, and finally, identify how we can fit the model to the data by
maximizing the likelihood.

\subsubsection{Notation for Tiles}
\label{sec:maxent:tile.notation}
Given a binary data matrix $D$ of size $N$-by-$M$ and a tile $T$, the frequency
of $T$ in $D$, $\mathit{fr}(T;D)$, is defined as
\begin{equation}
\label{eq:freq}
  \mathit{fr}(T;D) = \frac{1}{|\sigma(T)|} \sum\limits_{(i,j) \in \sigma(T)} D(i,j) \quad .
\end{equation}
Here, $D(i,j)$ represents the entry $(i,j)$ in $D$, and $\sigma(T) = \{(i,j)
\mid i \in r(T), j \in c(T)\}$ denotes the cells covered by tile $T$ in data
$D$. Recall that a tile $T$ is called `exact' if the corresponding entries
$D(i,j)$ $\forall (i,j) \in \sigma(T)$ are all $1$ (resp.\ $0$), or in other
words, $\mathit{fr}(T;D) = 0$ or $\mathit{fr}(T;D) = 1$. Otherwise, it is called
a `noisy' tile.

Let $\mathcal{D}$ be the space of all the possible binary data matrices
of size $N$-by-$M$, and $p$ be the probability distribution defined over the
data matrix space $\mathcal{D}$. Then, the frequency of the tile $T$ with
respect to $p$ is
\begin{equation}
\label{eq:expected.freq}
  \mathit{fr}(T;p) = \mathbb{E}\left[\mathit{fr}(T;D)\right]  =
  \sum\limits_{D \in \mathcal{D}} p(D)\mathit{fr}(T;D) \quad ,
\end{equation}
the expected frequency of tile $T$ under the data matrix probability distribution.

Combining these definitions, we have the following
lemma~\cite{Wu:2014:UPD:2664051.2664089}.
\begin{lemma}
\label{lemma:1}
  Given a dataset distribution $p$ and a tile $T$, the frequency of tile $T$ is
  \begin{equation*}
	\mathit{fr}(T;p) = \frac{1}{|\sigma(T)|} \sum\limits_{(i,j) \in \sigma(T)}
	p\left(\left(i,j\right) = 1\right) \quad ,
  \end{equation*}
  where $p((i,j) = 1)$ represents the probability of a data matrix having 1 at
  entry $(i,j)$ under the data matrix distribution $p$.
\end{lemma}
Lemma~\ref{lemma:1} is trivially proved by substituting $\mathit{fr}(T;D)$ in
Equation~\eqref{eq:expected.freq} with Equation~\eqref{eq:freq} and switching
the summations.

\subsubsection{Global MaxEnt Model from Tiles}
\label{sec:maxent:tile.global}
Here, we will construct a global statistical model based on tiles. Suppose we
are given a set of tiles $\mathcal{T}$, and each tile $T \in \mathcal{T}$ is
associated with a frequency $\gamma_{T}$---which typically can be trivially
obtained from the data. This tile set $\mathcal{T}$ provides information about
the data at hand, and we would like to infer a distribution $p$ over the space
of possible data matrices $\mathcal{D}$ that conforms with the information given
in $\mathcal{T}$. That is, we want to be able to determine how probable is a
data matrix $D \in \mathcal{D}$ given the tile set $\mathcal{T}$.

To derive a good statistical model, we take a principled approach and employ the
maximum entropy principle \cite{jaynes:59:maxent} from information theory.
Loosely speaking, the MaxEnt principle identifies the best distribution given
background knowledge as the unique distribution that represents the provided
background information but is maximally random otherwise. MaxEnt modeling has
recently become popular in data mining as a tool for identifying
\emph{subjective} interestingness of results with regard to background
knowledge~\cite{wang:06:summaxent,debie:11:dami,tatti:12:apples}.

To formally define a MaxEnt distribution, we first need to specify the
space of the probability distribution candidates. Here, these are all the
possible data matrix distributions that are consistent with the information
contained in the tile set $\mathcal{T}$. Hence, the data matrix distribution
space is defined as: $\mathcal{P} = \{p \mid \mathit{fr}(T;p) = \gamma_{T},
\forall T \in \mathcal{T}\}$. Among all these possible distributions, we choose
the distribution $p_{\mathcal{T}}^{*}$ that maximizes the entropy,
\begin{equation*}
  p_{\mathcal{T}}^{*} = \arg \max_{p \in \mathcal{P}} H(p) \quad .
\end{equation*}
Here, $H(p)$ represents the entropy of the data matrix probability distribution
$p$, which is defined as
\begin{equation*}
  H(p) = -\sum\limits_{D \in \mathcal{D}} p(D) \log p(D) \quad .
\end{equation*}
Next, to infer the MaxEnt distribution $p_{\mathcal{T}}^{*}$, we rely on a
classical theorem about how MaxEnt distributions can be factorized. In
particular, Theorem 3.1 in \cite{1975:Csiszar} states that for a given set of
testable statistics $\mathcal{T}$ (background knowledge, here a tile set), a
distribution $p_{\mathcal{T}}^{*}$ is the maximum entropy distribution if and
only if it can be written as
\begin{equation*}
  p_{\mathcal{T}}^{*}(D) \propto \left\{
  \begin{array}{ll}
    \exp\big( \sum\limits_{T \in \mathcal{T}} \lambda_{T} \cdot |\sigma(T)|
      \cdot \mathit{fr}(T;D)\big) & D \not\in \mathcal{Z} \\
      0 & D \in \mathcal{Z}\quad , \\
  \end{array} \right.
\end{equation*}
where $\lambda_{T}$ is a certain weight for $\mathit{fr}(T;D)$ and $\mathcal{Z}$
is a collection of data matrices such that $p(D) = 0$, for all $p \in
\mathcal{P}$.

\citet{debie:11:dami} formalized the MaxEnt model for a binary matrix $D$ given
row and column margins---also known as a \citet{rasch:60:probabilistic} model.
Here, we consider the more general scenario of binary data and tiles, for which
we additionally know \cite[Theorem 2 in][]{tatti:12:apples} that given a tile
set $\mathcal{T}$, with $\mathcal{T}(i,j) = \{T \in \mathcal{T} \mid (i,j) \in
\sigma(T)\}$, we can write the distribution $p_{\mathcal{T}}^{*}$ as
  \begin{equation*}
    p_{\mathcal{T}}^{*} = \prod\limits_{(i,j) \in D} p_{\mathcal{T}}^{*}((i,j) =
    D(i,j)) \quad ,
  \end{equation*}
where
  \begin{equation*}
    p_{\mathcal{T}}^{*}((i,j) = 1) = \frac{\exp\left(\sum_{T \in \mathcal{T}(i,j)}
        \lambda_{T}\right)}{\exp\left(\sum_{T \in \mathcal{T}(i,j)}
        \lambda_{T}\right) + 1} ~\text{or}~ 0,1 \quad .
  \end{equation*}
This result allows us to factorize the MaxEnt distribution $p_{\mathcal{T}}^{*}$
of binary data matrices given background information in the form of a set of
tiles $\mathcal{T}$ into a product of Bernoulli random variables, each of which
represents a single entry in the data matrix $D$. We should emphasize here that
this model is different MaxEnt model than when we assume independence between
rows in the data matrix $D$~\cite[see,
e.g.,][]{tatti:06:computational,wang:06:summaxent,mampaey:12:mtv}. Here, for
example, in the special case where the given tiles are all exact ($\gamma_{T} =
0$ or $1$), the resulting MaxEnt distribution will have a very simple form:
  \begin{equation*}
    p_{\mathcal{T}}^{*}((i,j) = 1) = \left\{
      \begin{array}{ll}
        \gamma_{T} & \text{if } \exists T \in \mathcal{T} \text{ such that }
        (i,j) \in \sigma(T)\\
        \frac{1}{2} & \text{otherwise.} \\
      \end{array} \right.
  \end{equation*}

\subsubsection{Inferring the MaxEnt Distribution}
\label{sec:maxent:iter.scaling}

 \begin{algorithm}[t]
  \SetAlgoLined
  \SetKwInOut{Input}{input}
  \SetKwInOut{Output}{output}

  \Input{a tile set $\mathcal{T}$, target frequencies
	$\{\gamma_{T}\mid T \in \mathcal{T} \}$.}
  \Output{maximum entropy distribution $p_{\mathcal{T}}^{*} \leftarrow p$.}
  \BlankLine
  $p \leftarrow$ a $N$-by-$M$ matrix with all values of $\frac{1}{2}$\;
  \For{$T \in \mathcal{T}, \gamma_{T} = 0, 1$}{
	$p(i,j) \leftarrow \gamma_{T}$, for all $(i,j) \in \sigma(T)$\;
  }
  \While{not converged}{
	\For{$T \in \mathcal{T}, 0 < \gamma_{T} < 1$}{
	  find $x$ such that: $\mathit{fr}(T;p) = \sum_{(i,j) \in \sigma(T)}
	  \frac{x \cdot p(i,j)}{1 - (1 - x) \cdot p(i,j)}$\;
	  $p(i,j) \leftarrow \frac{x \cdot p(i,j)}{1 - (1 - x) \cdot p(i,j)}$, for all $(i,j) \in \sigma(T)$\;
	}
  }
  \caption{Iterative Scaling Algorithm (binary dataset)} \label{alg:1}
 \end{algorithm}

To discover the parameters of the Bernoulli random variable mentioned above, we
follow a standard approach and apply the well known Iterative Scaling (IS)
algorithm \cite{1972:iterative:scaling} to infer the tile based MaxEnt
distribution on binary data matrices. Algorithm~\ref{alg:1} illustrates the
details of this IS algorithm for binary data. Basically, for each tile $T
\in \mathcal{T}$, the algorithm updates the probability distribution $p$ such
that the expected frequency of $1$s under the distribution $p$ will match
the given frequency $\gamma_{T}$. Clearly, during this update we may change the
expected frequency for other tiles, and hence several iterations are needed
until the probability distribution $p$ converges. For a proof of convergence,
please refer to Theorem 3.2 in \cite{1975:Csiszar}. In practice, the algorithm
typically takes on the order of seconds to converge.

\subsection{Maximum Entropy Model for Real-valued Data}
\label{sec:real_maxent}
In this section, we introduce the MaxEnt model for real-valued data with tiles
as background knowledge. We first extend the concept of tiles from binary
transactional matrix to a real-valued transactional matrix. Then, we formulate
the global MaxEnt model over the real-valued transactional data, and finally, we
provide an efficient algorithm to infer the real-valued MaxEnt distribution.

\subsubsection{Notation for Tiles}
\label{sec:real_tiles}
As stated earlier, a document-entity transactional matrix $D$ usually contains
occurrence (count) information for each entity in every document of the corpus. 
Count data is integer valued but without loss
of generality, the entries in the real-valued transactional matrix $D$ is
considered to be normalized into the range of $[0,1]$ (e.g.\ each entry of $D$
can be divided by the maximum entry of $D$).

A tile $T$ over a real-valued matrix $D$ is still defined as the tuple $T =
(r(T), c(T))$ which identifies a sub-matrix from $D$. Compared to the frequency
of a tile defined in the binary case, more descriptive statistical measures can
be defined for real-valued tiles. In our scenario, we choose the sum of the
values and sum of the squared values identified by a tile $T$, which are
represented by $f_m$ and $f_v$ respectively. More specifically, $f_m$ and $f_v$
are defined as follow:
\begin{align}
	\label{eq:def_mean_var}
	f_m(T \mid D) & = \sum_{\forall (i,j) \in \sigma(T)} D(i,j) \\
	f_v(T \mid D) & = \sum_{\forall (i,j) \in \sigma(T)} D^2(i,j) \nonumber
\end{align}

\subsubsection{Global MaxEnt Model from Tiles}
\label{sec:maxent_real_tile}
A real-valued MaxEnt model was first proposed
by~\citet{real:value:maxent}. Given a set of real-valued tiles $\mathcal{T}$
where for every entry $(i,j)$ in the matrix $D$, there exists at least a tile $T
\in \mathcal{T}$ such that $(i,j) \in \sigma(T)$. Each tile $T \in \mathcal{T}$
is associated with $\tilde{f}_m(T)$ and $\tilde{f}_v(T)$ as its basic
statistics. Then, the probability distribution space of real-valued data
matrices can be defined as 
$$
\mathcal{P} = \{p \mid \mathbb{E}_{p}[f_m(T \mid D)] = \tilde{f}_m(T),
\mathbb{E}_p[f_v(T \mid D)] = \tilde{f}_v(T), \forall T \in \mathcal{T} \}~.
$$
Here, $\tilde{f}_m$ and $\tilde{f}_v$ denote the empirical values of the
statistics associated with tiles, which can be computed from the given
real-valued data matrix, and $\mathbb{E}_{p}[\cdot]$ represents the expectation
with respect to the probability distribution $p$. Among all the candidate
distribution $p \in \mathcal{P}$, we choose the one that maximizes the entropy
according to:
$$
p^{*}_{\mathcal{T}} = \argmax_{p \in \mathcal{P}} \left\{ -\oint\limits_{D} p(D)
\log p(D) dD \right\}~.
$$

To be more specific, inferring the MaxEnt distribution could be formulated as
the following optimization problem:
\begin{align}
	\label{eq:maxent_optimize}
	p^{*}_{\mathcal{T}} & = \argmax_{p} \left\{ -\oint\limits_{D} p(D) \log p(D)
	dD \right\} \\
	\text{s.t.} \quad & \oint\limits_{D} p(D) f_{m}(T \mid D) d D =
	\tilde{f}_{m}(T),~\forall T \in \mathcal{T} \nonumber \\
	& \oint\limits_{D} p(D) f_{v}(T \mid D) d D = \tilde{f}_{v}(T),~\forall T \in
	\mathcal{T} \nonumber \\
	& \oint\limits_{D} p(D) d D = 1,~p(D) \geq 0 \nonumber
\end{align}
Since the optimization problem defined above is convex, by applying the approach
of Lagrange multipliers, we can derive that the MaxEnt distribution has the
following exponential form:
$$
p^{*}_{\mathcal{T}}(D) = \frac{1}{Z} \exp\left(-\sum_{T \in \mathcal{T}}
\lambda^{(m)}_{T} f_m(T \mid D) - \sum_{T \in \mathcal{T}} \lambda^{(v)}_{T}
f_v(T \mid D) \right)~.
$$
Substituting $f_m(T \mid D)$ and $f_v(T \mid D)$ with their definitions from
Equation~\eqref{eq:def_mean_var}, the MaxEnt distribution could be simplified
as:
\begin{align}
	\label{eq:maxent_factorized}
	p^{*}_{\mathcal{T}} & = \frac{1}{Z} \prod_{(i,j) \in D} \exp
	\left(-\beta_{i,j} D^{2}(i,j) - \alpha_{i,j} D(i,j) \right) \\
	& = \prod_{(i,j) \in D} p_{i,j}(D(i,j)) \nonumber
\end{align}
where
\begin{align*}
	p_{i,j}(D(i,j)) & = \sqrt{\frac{\beta_{i,j}}{\pi}} \exp\left\{ -\frac{\left(
		D(i,j) + \frac{\alpha_{i,j}}{2\beta_{i,j}}\right)^2}{1 / \beta_{i,j}}
	\right\} \\ 
	\alpha_{i,j} & = \sum_{\substack{(i,j) \in \sigma(T) \\ T \in \mathcal{T}}}
	\lambda_{T}^{(m)}, \quad 
	\beta_{i,j} = \sum_{\substack{(i,j) \in \sigma(T) \\ T in \mathcal{T}}}
	\lambda_{T}^{(v)} \nonumber
\end{align*}
Equation~\eqref{eq:maxent_factorized} indicates that the real-valued MaxEnt
distribution over the matrix $D$ could be factorized into the product of the
distributions of $D(i,j)$ where each $D(i,j)$ follows the Gaussian distribution:
$$
D(i,j) \sim \mathbb{N}\left( -\frac{\alpha_{i,j}}{2 \beta_{i,j}},
\frac{1}{2 \beta_{i,j}}\right)
$$
In addition, we can also compute the normalizing constant $Z$ in
Equation~\eqref{eq:maxent_factorized} as
\begin{align*}
	Z & = \oint\limits_{D} \prod_{(i,j) \in D} \exp \left(-\beta_{i,j} D^2(i,j)
	- \alpha_{i,j} D(i,j) \right) d D \\
	& = \prod_{(i,j) \in D} \sqrt{\frac{\pi}{\beta_{i,j}}} \exp \left(
	\frac{\alpha^{2}_{i,j}}{4 \beta_{i,j}}\right)
\end{align*}

\subsubsection{Inferring the MaxEnt Distribution}
\label{sec:infer_real}

\begin{algorithm}[t]
  \SetAlgoLined
  \SetKwInOut{Input}{input}
  \SetKwInOut{Output}{output}
  \SetKwFunction{updateAlphaBeta}{updateAlphaBeta}

  \Input{a tile set $\mathcal{T}$, target tile statistics
  $\{f_{m}(T \mid D), f_{v}(T \mid D) \mid T \in \mathcal{T} \}$.}
  \Output{Maximum Entropy distribution $p_{\mathcal{T}}^{*}$ parameterized by
  $\alpha_{i,j}$ and $\beta_{i,j}$.}
  \BlankLine

  Initialize $\lambda_{T}^{(m)}$ and $\lambda_{T}^{(v)}$ randomly $\forall T \in
  \mathcal{T}$\;
  $\boldsymbol{\lambda} \leftarrow [\lambda_T^{(m)}, \lambda_T^{(v)} \mid T \in
  \mathcal{T}]$\;
  \While{not converged}{
	  \updateAlphaBeta{$\boldsymbol{\lambda}$}\;
	  compute gradient using Equation~\eqref{eq:grad_m} and~\eqref{eq:grad_v}\;
	  perform a conjugate gradient update on $\boldsymbol{\lambda}$\;
  }
  \caption{MaxEnt model inference (real-valued dataset)} \label{alg:2}
\end{algorithm}

To infer the real-valued MaxEnt distribution, we need to estimate the
values of the model parameters $\lambda_{T}^{(m)}$ and $\lambda_{T}^{(v)}$. 
We leverage the duality between maximum entropy and maximum likelihood
formulations~\cite{maxent-duality} by solving the following problem:
\begin{align*}
	\max_{\boldsymbol{\lambda}}: \mathcal{L}(\boldsymbol{\lambda}) & = \log p(D)
	= \sum_{T \in \mathcal{T}} \left( -\lambda_{T}^{(m)} \tilde{f}_{m}(T) -
	\lambda_{T}^{(v)} \tilde{f}_{v}(T) \right) - \log Z \\
	& = - \sum_{(i,j) \in D} \left[ \frac{1}{2} \log \left(
		\frac{\pi}{\beta_{i,j}}\right) +
	\frac{\alpha_{i,j}^{2}}{4\beta_{i,j}}\right] + \sum_{T \in \mathcal{T}}
	\left( -\lambda_{T}^{(m)} \tilde{f}_{m}(T) - \lambda_{T}^{(v)}
	\tilde{f}_{v}(T) \right) \\
	\text{s.t.} \quad & \beta_{i,j} > 0, \quad \forall (i,j) \in D
\end{align*}
The above optimization problem is convex and can be solved
efficiently by state-of-the-art optimization algorithms. Here, we choose the
conjugate gradient method to solve this problem, where the gradient of the
objective function $\mathcal{L}(\boldsymbol{\lambda})$ is given by:
\begin{align}
	\frac{\partial \mathcal{L}(\boldsymbol{\lambda})}{\partial
		\lambda_{T}^{(m)}} & = - \sum_{(i,j) \in \sigma(T)} \left(
		\frac{\alpha_{i,j}}{2 \beta_{i,j}} \right) - \tilde{f}_{m}(T)
		\label{eq:grad_m} \\
	\frac{\partial \mathcal{L}(\boldsymbol{\lambda})}{\partial
		\lambda_{T}^{(v)}} & = \sum_{(i,j) \in \sigma(T)} \left(
		\frac{1}{2\beta_{i,j}} + \frac{\alpha_{i,j}^{2}}{4 \beta_{i,j}^{2}}
		\right) - \tilde{f}_{v}(T) \label{eq:grad_v}
\end{align}

\section{Scoring Biclusters and Chains}
\label{sec:score}
We now turn our attention to using the above formalisms to help score our
patterns, viz., biclusters and bicluster chains. But before we do so, we need to
pay attention to the relational schema over which these patterns are inferred,
as this influences how patterns can be represented as tiles, in order to be
incorporated as knowledge in our maximum entropy models.

\subsection{Entity-Entity Relation Extraction}
\label{sec:data_model}
In this section, we describe the approach to construct a multi-relational schema
$\schema(\mathcal{U}, \mathcal{R})$ from a transaction data matrix $D$. Recall
that whenever an element $D(r, e_i)$ has a non-zero value (e.g.\ 1 in the binary
case or a fraction in the range of $[0,1]$ in the real-valued case), this
denotes that entity $e_i$ appears in row $r$ of $D$. As an example, when
considering text data, an entity would correspond to a word or concept, and a
row to a document in which this word occurs. (Thus, note that when considering
text data we currently model occurrences of entities at the granularity of
documents. Admittedly, this is a coarse modeling in contrast to modeling
occurrences at the level of sentences, but it suffices for our purposes.)



To extract entity-entity relations from transaction data matrix $D$, we utilize
the entity co-occurrence information. To be more specific, each binary relation
in $\mathcal{R}$ stores the entity co-occurrences in data matrix $D$ between two
entity domains, e.g.\ for each $R = R(U_i, U_j)$ in $\mathcal{R}$, $(e, f) \in
R$ for $e \in U_i$, $f \in U_j$, and $e$ and $f$ appear at least once together
in a row in $D$. 


\subsection{Background Model Definition}
\label{sec:back_model}
Next, to discover non-trivial and interesting patterns, we need to incorporate
some basic information about the multi-relational schema $S(\mathcal{U},
\mathcal{R})$ into the model. For such basic background knowledge over $D$ we
use the column marginals and the row marginals for each entity domain. To this
end, following~\citet{Wu:2014:UPD:2664051.2664089} we construct a tile set
$\mathcal{T}_\mathit{col}$ consisting of a tile per column, a tile set
$\mathcal{T}_\mathit{row}$ consisting of a tile per row per entity domain, and a
tile set $\mathcal{T}_\mathit{dom}$ consisting of a tile per entity domain but
spanning all rows. Formally, we have
\begin{eqnarray*}
    \mathcal{T}_\mathit{col} & = & \{(U_D, e) \mid  e \in U, U \in
    \mathcal{U} \setminus \{U_D\} \}, \\
    \mathcal{T}_\mathit{row} & = & \{(r, U) \mid r \in U_D, U \in
    \mathcal{U} \setminus \{U_D\} \}, \text{ and}  \\
	\mathcal{T}_\mathit{dom} & = & \{(U_D, U) \mid U \in \mathcal{U}
	\setminus \{U_D\} \}.
\end{eqnarray*}
Here, $U_{D}$ represents the domain of all the documents in the dataset. We
refer to the combination of these three tile sets as the background tile set
$\mathcal{T}_\mathit{back} = \mathcal{T}_\mathit{row} \cup
\mathcal{T}_\mathit{col} \cup \mathcal{T}_\mathit{dom}$. Given the background
tiles $\mathcal{T}_\mathit{back}$, the background MaxEnt model $p_\mathit{back}$
can be inferred using iterative scaling (see
Sect.~\ref{sec:maxent:iter.scaling}) and the conjugate gradient method (see
Sect.~\ref{sec:infer_real}) for binary and real-valued cases, respectively.


\subsection{Quality Scores}
\label{sec:quality_score}
To assess the quality of a given bicluster $B$ with regard to our background
knowledge, we need to first convert it into tiles such that we can infer the
corresponding MaxEnt model. Below we specify how we do this conversion
for biclusters from entity-entity relations. For a given bicluster $B = (E_i,
E_j)$, we construct a tile set $\mathcal{T}_B$, consisting of
$\abs{E_i}\abs{E_j}$ tiles, as follows
\begin{equation}
  \label{eq:biTiles}
  \mathcal{T}_B = \{\left(\rows(X; D),\  X \right) \mid X = \{e_i,
  e_j\} \mathrm{~with~} (e_i, e_j) \in B\}\quad, 
\end{equation}
where $\rows(X; D)$ is the set of rows that contain $X$ in $D$, e.g.\ the
corresponding entries for $X$ in the matrix $D$ that have non-zero values. 


To evaluate the quality of a bicluster chain $C$, for each bicluster $B \in C$,
we construct the set of tiles $\mathcal{T}_{B}$ as illustrated by
Equation~\eqref{eq:biTiles}, and the tile set that corresponds to a bicluster
chain $C$ is then $\mathcal{T}_\mathit{C} = \bigcup_{B \in \mathit{C}}
\mathcal{T}_{B}$.



Next, we describe the metrics that measure how much information a bicluster
$B$ (or the corresponding tile set $\mathcal{T}_{B}$) gives with regard to the
background model $p_{\mathit{back}}$. The global score is defined as follows:
\begin{align}
	s_{\mathit{global}}(B) = \mathit{KL}(p_{B} || p_{\mathit{back}})~,
	\label{eq:global}
\end{align}
where $p_{B}$ represents the MaxEnt distribution inferred over the background
tile set $\mathcal{T}_{\mathit{back}}$ and the tile set $\mathcal{T}_{B}$ for
the bicluster $B$. 

For both of binary and real-valued MaxEnt model, the MaxEnt distribution $p(D)$
can be factorized as 
$$
p(D) = \prod_{(i,j) \in D} p(D(i,j))~.$$
Thus, this global score can be written as:
\begin{align}
	s_{\mathit{global}}(B) & = \oint\limits_{D} p_{B}(D) \log
	\frac{p_{B}(D)}{p_{\mathit{back}}(D)} d D \nonumber \\
	& = \oint\limits_{D} \prod_{(i,j) \in D} p_{B}(D(i,j)) \sum_{(i,j) \in D}
	\log \frac{p_{B}(D(i,j))}{p_{\mathit{back}} (D(i,j))} d D \nonumber \\
	& = \sum_{(i,j) \in D} \int_{-\infty}^{+\infty} p_{B}(D_{{i,j}}) \log
	\frac{p_{B}(D(i,j))}{p_{\mathit{back}}(D(i,j))} d D(i,j) \nonumber \\
	& = \sum_{(i,j) \in D} \mathit{KL}(p_{B}(D(i,j)) ||
	p_{\mathit{back}}(D(i,j)))~.
	\label{eq:global_real}
\end{align}
For the binary MaxEnt model, $D(i,j)$ follows the Bernoulli distribution
$$
D(i,j) \sim \mathit{Bernoulli}(q),~\text{where}~q = \frac{\exp\left(\sum_{T
	\in \mathcal{T}(i,j)} \lambda_{T}\right)}{\exp\left(\sum_{T \in
	\mathcal{T}(i,j)} \lambda_{T}\right) + 1}~,
$$
and the global score for binary MaxEnt model would be:
$$
s_{\mathit{global}}(B) = \sum_{(i,j) \in D} \left(q_{B} \log
\frac{q_B}{q_{\mathit{back}}} + (1 - q_B) \log \frac{1 - q_B}{1 -
	q_{\mathit{back}}}\right)~.
$$
For the real-valued MaxEnt model, $D(i,j)$ follows the Gaussian distribution
$$
D(i,j) \sim \mathbb{N}\left(-\frac{\alpha_{i,j}}{2 \beta_{i,j}}, \frac{1}{2
\beta_{i,j}}\right)~.
$$
Given any two normal distribution $P_{\mathcal{N}_1} = \mathcal{N}(\mu_1,
\sigma_1^2)$ and $P_{\mathcal{N}_2} = \mathcal{N}(\mu_2, \sigma_2^2)$, we can
verify that the KL-divergence between these two normal distribution is:
\begin{align}
	\mathit{KL}(P_{\mathcal{N}_1} || P_{\mathcal{N}_2}) = \log
	\frac{\sigma_2}{\sigma_1} + \frac{\sigma_1^2 + {(\mu_1 - \mu_2)}^2}{2
	\sigma_2^2} - \frac{1}{2}~.
	\label{eq:kl_normal}
\end{align}
Combining Equation~\eqref{eq:global_real} and~\eqref{eq:kl_normal}, the global
score for the real-valued maximum entropy model is:
\begin{align}
	s_{\mathit{global}} = \sum_{(i,j) \in D} \left[ \frac{1}{2} \log
		\frac{\beta_{i,j}^{(B)}}{\beta_{i,j}^{(\mathit{back})}} +
		\frac{\beta_{i,j}^{(\mathit{back})}}{2 \beta_{i,j}^{(B)}} +
		\beta_{i,j}^{(\mathit{back})} {\left(
		\frac{\alpha_{i,j}^{(\mathit{back})}}{2
		\beta_{i,j}^{(\mathit{back})}} - \frac{\alpha_{i,j}^{(B)}}{2
		\beta_{i,j}^{(B)}}\right)}^2 - \frac{1}{2} \right]
	\label{eq:global_real_final}
\end{align}

However, using the global score defined above requires us to re-infer the MaxEnt model
for every candidate bicluster that needs to be evaluated, which could be
computationally expensive and thus not applicable to our
interactive mining sitting. Moreover, $s_{\mathit{global}}$ evaluates a candidate
globally, whereas typically most information is \textit{local}: at most a few entries
in the maximum entropy distribution will be affected by adding $B$ into the model.
Making use of this observation, to reduce the computational cost of candidate
bicluster evaluation, we define the score $s_{\mathit{local}}(B)$ that measures
the local surprisingness of a tile set as
\begin{align}
	s_{\mathit{local}}(B) = - \sum_{T \in \mathit{T}_{B}} \sum_{(i,j) \in
	\sigma(T)} \log p_{\mathit{back}}(D(i,j))~,
	\label{eq:local}
\end{align}
where for both binary and real-valued MaxEnt model, $p_{\mathit{back}}(D(i,j))$
indicates the probability (or probability density) evaluated at the value
$D(i,j)$ under the current background MaxEnt model. Notice that although the
global and local scores are described using the notation of biclusters here,
they can also be directly adopted to assess the quality of bicluster chains
because fundamentally these scores are defined around the concept of tiles and
bicluster chains (and can thus be trivially converted to a set of tiles as described at
the beginning of this section).

\section{\MaxEntBiSet}
\label{sec:biset}

\MaxEntBiSet is a visual analytics system, supported by the 
maximum entropy model above,
to support interactive exploration of coordinated relationships
using biclusters. Coordinated relationships are groups of relations, connecting
sets of entities from different domains (e.g., people, location, organization,
etc.), which potentially indicate coalitions between these entities.  \MaxEntBiSet
extends an existing bicluster visualization, viz. BiSet~\cite{sun2015biset}, by
incorporating MaxEnt models to support user exploration of entity coalitions
for sensemaking purposes. In this section, we first introduce BiSet, followed
by the enhancements that
\MaxEntBiSet provides.

\subsection{BiSet Technique Overview}

\begin{figure}[!t]
  \centering
  \includegraphics[width=0.9\textwidth]{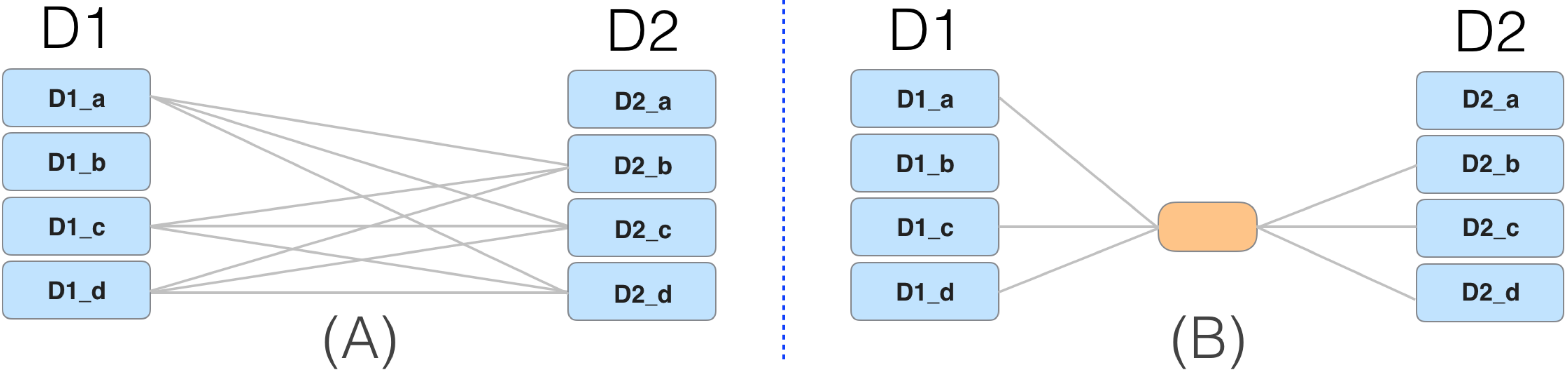}
  \caption{Visual representations of a bicluster that includes three entities in \textsl{D1}
  and three entities in \textsl{D2}. (A) displays all individual relationships between
  the two sets of entities from the two domains, \textsl{D1} and \textsl{D2}.
  (B) Realtionships are aggregated as an edge bundle that
  represents a bicluster.}
  \label{fig:biset-bic}
\end{figure}

\begin{figure}[!t]
  \centering
  \includegraphics[width=0.9\textwidth]{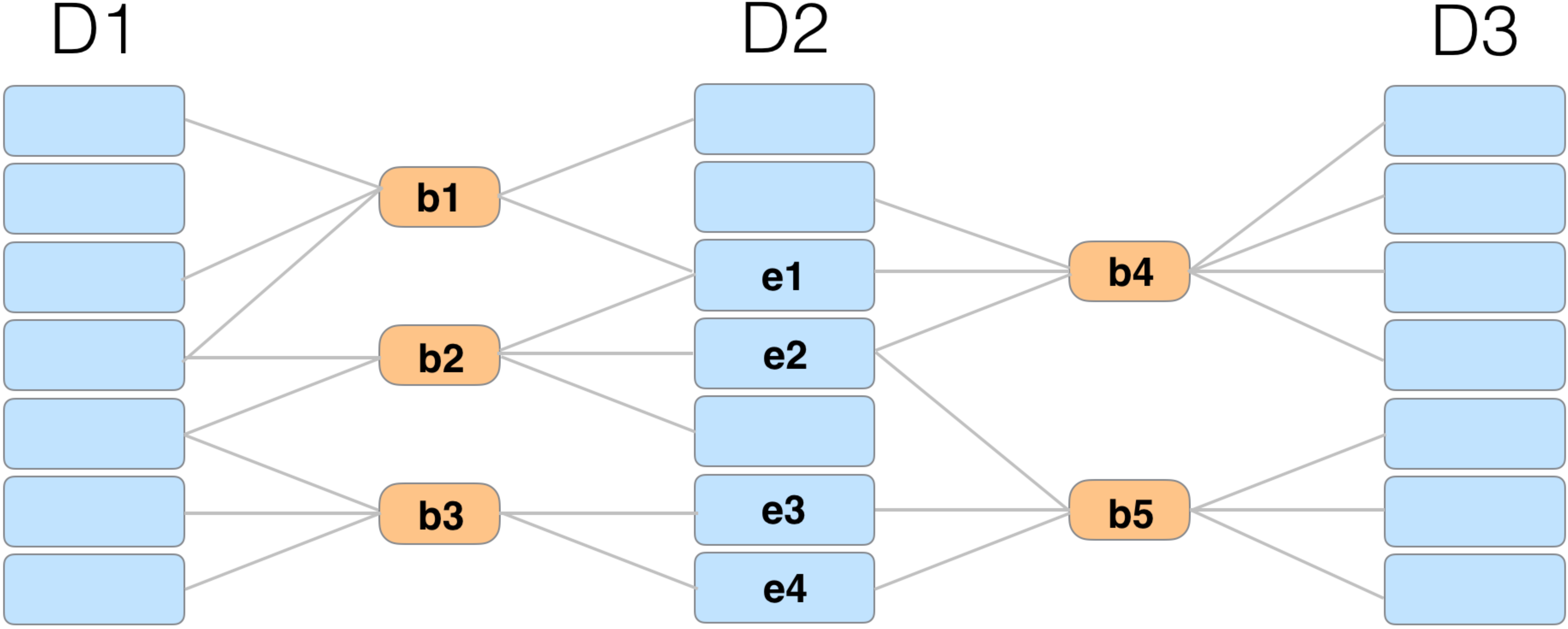}
  \caption{An example of four bicluster-chains (\textsl{b1} - \textsl{b4},
  \textsl{b2} - \textsl{b4}, \textsl{b2} - \textsl{b5} and \textsl{b3} -
  \textsl{b5}). These chains consist of entities from three domains,
  \textsl{D1}, \textsl{D2} and \textsl{D3}. \textsl{b1} and \textsl{b4} are
  connected through \textsl{e1}. \textsl{b2} and \textsl{b4} share \textsl{e1}
  and \textsl{e2}. \textsl{b2} and \textsl{b5} are linked by \textsl{e3}.
  \textsl{b3} and \textsl{b5} are connected by \textsl{e3} and \textsl{e5}.}
  \label{fig:biset-chain}
\end{figure}

The key idea is
that \textsl{BiSet visualizes the mined biclusters in
context as edge bundles} between sets of related entities. BiSet uses lists as
the basic layout to present entities and biclusters. Figure~\ref{fig:biset-bic}
shows an example of a visualized bicluster in BiSet. In
Figure~\ref{fig:biset-bic}, (A) shows all individual edges between related
entities and (B) presents the same bicluster as an edge bundle. BiSet enables
both ways to show the coalition of entities with two modes: \textsl{link mode}
and \textsl{bicluster mode}. \textsl{Link mode} displays the individual
connections among entities in a dataset, while \textsl{bicluster mode} offers a
more clear representation to show identified biclusters in the dataset. Based on
these visual representations, BiSet can visually show bicluster-chains as
connected edge bundles through their shared entities.
Figure~\ref{fig:biset-chain} shows four bicluster-chains (\textsl{b1} -
\textsl{b4}, \textsl{b2} - \textsl{b4}, \textsl{b2} - \textsl{b5} and
\textsl{b3} - \textsl{b5}) visualized using BiSet.  Each of them consists of two
different biclusters including entities from three domains. The two biclusters
in each chain are visually connected through one or two shared entities. For
example, bicluster \textsl{b2} and \textsl{b4} are connected by entity
\textsl{e1} and \textsl{e2}. With edges, BiSet enables users to see members of
bicluster-chains and how these biclusters are connected. This potentially guides
users to interpret the coalition among sets of entities from multiple domains in
an organized manner (e.g., checking connected biclusters from left to right).

To support exploratory analysis, BiSet treats \textsl{edge bundles as first class
objects}, so users can directly manipulate them (e.g., drag and move) to
spatially organize them in meaningful ways. BiSet also offers automatic
ordering for entities and biclusters to help users organize them. For example,
entities can be ordered based on their frequency in a dataset and biclusters can
be ordered by size, i.e., the number of entities participating in a bicluster.
Moreover, BiSet can highlight bicluster-chains as users select their members.
This provides visual clues for users to follow in conducting their analysis.

\subsection{\MaxEntBiSet Visual Encoding}

\begin{figure}[!t]
  \centering
  \includegraphics[width=1\textwidth]{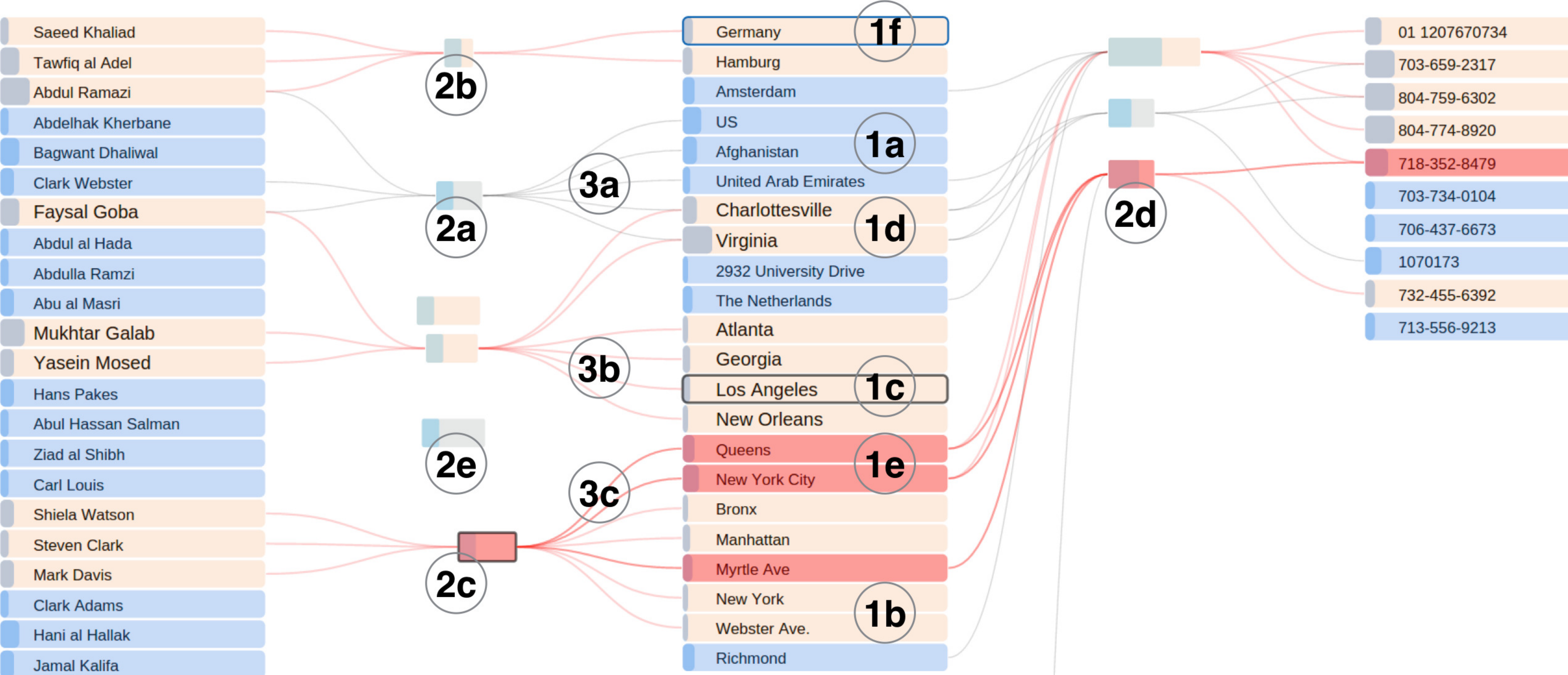}
  \caption{Detailed visual encodings in \MaxEntBiSet. \textsl{1a}, \textsl{2a} and
  \textsl{3a} depict the \textsl{normal} state of an entity, a bicluster and
  edges, respectively. \textsl{1b}, \textsl{2b} and \textsl{3b} depict the
  \textsl{connection-oriented} highlighting state of an entity, a bicluster and
  edges, when users \textsl{select} bicluster \textsl{2c}, \textsl{hover} over entity \textsl{1f}
  and \textsl{select} entity \textsl{1c}.
  \textsl{1e}, \textsl{2d} and \textsl{3c} illustrate the 
\textsl{surprisingness-oriented} highlighting state of an entity, a bicluster and edges. \textsl{1d}
  demonstrates larger fonts of entities as users hover the mouse pointer over
  their previously selected entity \textsl{1c}. Moreover, \textsl{2e} represents a
  bicluster (in the \textsl{normal} state) with its edges chosen to be hidden by users.}
  \label{fig:biset-visual}
\end{figure}

\subsubsection{Shape and Size}

In \MaxEntBiSet, entities and biclusters are represented as rectangles (e.g.,
\textsl{1a} and \textsl{2a} in Figure~\ref{fig:biset-visual}), and edges are
visualized as B\'ezier curves. We use B\'ezier curves because they can generate
more smooth edges, compared with polylines~\cite{Lambert:2010jg}. Rectangles
indicating entities are equal in length, while those representing biclusters are
not. \MaxEntBiSet applies a linear mapping function to determine the length of a bundle
based on the total number of its related entities. In a bicluster rectangle,
\MaxEntBiSet uses two colored regions (light green and light gray) to indicate the
proportion between its related entities in lists of both sides (left and right).
In an entity rectangle, a small rectangle is displayed on the left to indicate
its frequency in a dataset. The length of these rectangles is determined by the
frequency of the associated entities with a linear mapping function. These helps
users to visually discriminate entities from biclusters. Moreover, when users
hover over a selected entity or bicluster (e.g., entity \textsl{1c} and bicluster
\textsl{2c} in Figure~\ref{fig:biset-visual}), the font of its related entities
is enlarged (e.g., comparing \textsl{1d} with \textsl{1b} in
Figure~\ref{fig:biset-visual}). This helps users review relevant information
of their previous selections.

\subsubsection{Color Coding}

\MaxEntBiSet applies color coding to entities, biclusters and edges to indicate their
states and allows users to hide edges of biclusters to reduce visual clutter 
(see \textsl{2e} in Figure \ref{fig:biset-visual}). 
In \MaxEntBiSet, entities, biclusters and edges have two basic states:
\textsl{normal} and \textsl{highlighted}. The normal state is the default state
for entities, biclusters and edges. Examples of the normal state for them are
shown as \textsl{1a}, \textsl{2a} and \textsl{3a}, respectively, in
Figure~\ref{fig:biset-visual}. To encode surprisingness, 
\MaxEntBiSet supports two types of highlighting states:
\textsl{connection oriented highlighting} (colored as \textsl{orange} in
Figure~\ref{fig:biset-visual}) and \textsl{surprisingness oriented highlighting}
(color as \textsl{red} in Figure~\ref{fig:biset-visual}), which encode two
levels of information: \textsl{the coalition of entities} and \textsl{the
surprisingness of the coalition}. The former indicates the linkage of entities,
emphasizing the connections between entities. The latter reveals the 
model-evaluated surprisingness of different sets of entity coalitions. In
Figure~\ref{fig:biset-visual}, examples of connection-oriented highlighting
for entities, biclusters and edges are shown as \textsl{1b}, \textsl{2b}
and \textsl{3b}, respectively; while examples of surprisingness-oriented
highlighting are presented as \textsl{1e}, \textsl{2d} and
\textsl{3c}. 

The connection oriented highlighting state is triggered as users hover or select
an entity or a bicluster. For example, when users hover the mouse pointer over the entity \textsl{1f},
its directly connected bicluster \textsl{2b} is highlighted and other entities
that belong to this bicluster are also highlighted. The surprisingness oriented
highlighting state is triggered by explicit user request of model evaluation.
For instance, in Figure~\ref{fig:biset-visual}, as users request to find the
most surprising chains with bicluster \textsl{2c} as the starting point,
\MaxEntBiSet highlights entities and biclusters in a chain that has the highest
score given by the proposed \textsl{Maximum Entropy model} (the approach to
discover such a chain will be described in Section~\ref{biset-model} below).
With our color codings, \MaxEntBiSet 
empowers users to explore entity coalitions by directing them to computationally
identified surprising chains.

\subsection{Human-model Interaction}

\begin{figure}[!t]
  \centering
  \includegraphics[width=1\textwidth]{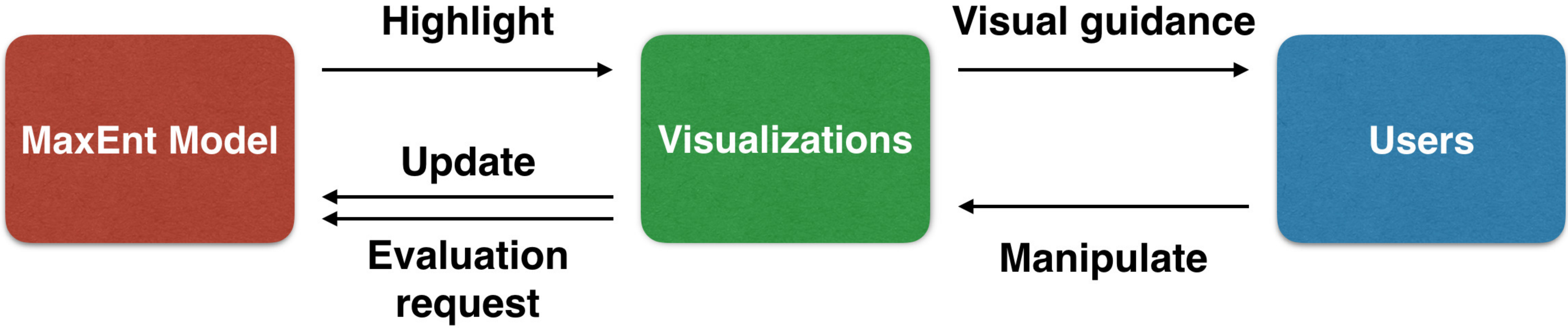}
  \caption{The human-model interaction flow in \MaxEntBiSet. Visual
	  representations in \MaxEntBiSet enable the interaction between users and
	  the proposed maximum entropy models.}\label{fig:biset-human-model}
\end{figure}

\MaxEntBiSet allows human-model interaction with visualizations to support
visual analytics of entity coalitions. To enable this capability, we incorporate
the proposed \textsl{Maximum Entropy models} into \MaxEntBiSet.
Figure~\ref{fig:biset-human-model} illustrates the human-model interaction flow
in \MaxEntBiSet. Visual representations in \MaxEntBiSet work as the bridge to
enable the interaction between users and the proposed models. 
After inspecting the
visualized biclusters and bicluster-chains, users can explicitly request model
evaluations using right click menus on a bicluster. This further triggers the
maximum entropy model to evaluate either all paths passing through the
requested bicluster or its neighboring biclusters. Then, based on results of the
model evaluation, \MaxEntBiSet highlights the most surprising bicluster-chain
including the user requested bicluster or neighboring biclusters. We address
this with a detailed discussion in Section~\ref{biset-model}. Moreover, users
can mark highlighted bicluster(s), based on model evaluation, as useful one(s)
by using a right click menu on the bicluster(s). This implicitly evokes a model
update function, which informs the model that the information in a marked
bicluster has been known by users. Then the model updates its background
information to take the marked bicluster(s) into account and prepare for
further user requested evaluations. This human-model interaction flow in
\MaxEntBiSet enables the combination the human cognition with computations for the
exploration of entity coalitions.

\subsection{Model Evaluation Strategies}
\label{biset-model}
\MaxEntBiSet offers two strategies to evaluate bicluster-chains, using the proposed
maximum entropy models, based on explicit user requests: \textsl{full
path evaluation} and \textsl{stepwise evaluation}. Both ways require users to
explicitly specify a bicluster based on its visual information, e.g.\ size
of a bicluster, frequency of corresponding entities, etc., to initiate the
chain.  The former evaluates all bicluster-chains passing through the bicluster
that users request for evaluation, while the latter evaluates neighboring
biclusters that satisfy a certain degree of overlap with the user-specified one.
\MaxEntBiSet enables users to explicitly issue an evaluation request from a
bicluster with a right click menu. From the menu, users can choose the desired
way of evaluation.

\subsubsection{Full Path Evaluation}
 
The \textsl{full path evaluation} in \MaxEntBiSet includes three key steps: 1) path
search, 2) path evaluation, and 3) path rank. In \MaxEntBiSet, a path, passing through
a bicluster, refers to a set of biclusters (e.g., \{\textsl{b2}, \textsl{b4}\}
in Figure~\ref{fig:biset-chain}), which can be connected through certain
entities to form a bicluster-chain. In the \textsl{path search} step, \MaxEntBiSet
finds all possible paths passing through the bicluster that users request for
evaluation. Similar to tree search, \MaxEntBiSet treats the user requested bicluster as
a root node and applies depth-first search to find all paths starting from this
bicluster. If the user requested bicluster is not from the left or right
most relation in the user specified multi-relational schema, \MaxEntBiSet
performs bidirectional search and then combines identified paths in the left and
those in the right together to obtain all paths going through this bicluster. Then
in the \textsl{path evaluation} step, \MaxEntBiSet converts each
bicluster-chain, found in the previous step, into a unique set of tiles following
the Equation~\eqref{eq:biTiles} in Section~\ref{sec:quality_score}, and applies
the maximum entropy models to score them. Finally, based on the score
from the model, in the \textsl{path rank} step, \MaxEntBiSet ranks these
bicluster-chains and visually highlights the one that has the highest score
(e.g., \{\textsl{2c}, \textsl{2d}\} in Figure~\ref{fig:biset-visual}).  Thus,
with the \textsl{full path evaluation} in \MaxEntBiSet, users can get the most
surprising bicluster-chain for the bicluster requested for evaluation.

\subsubsection{Stepwise Evaluation}

\begin{figure}[!t]
  \centering
  \includegraphics[width=1\textwidth]{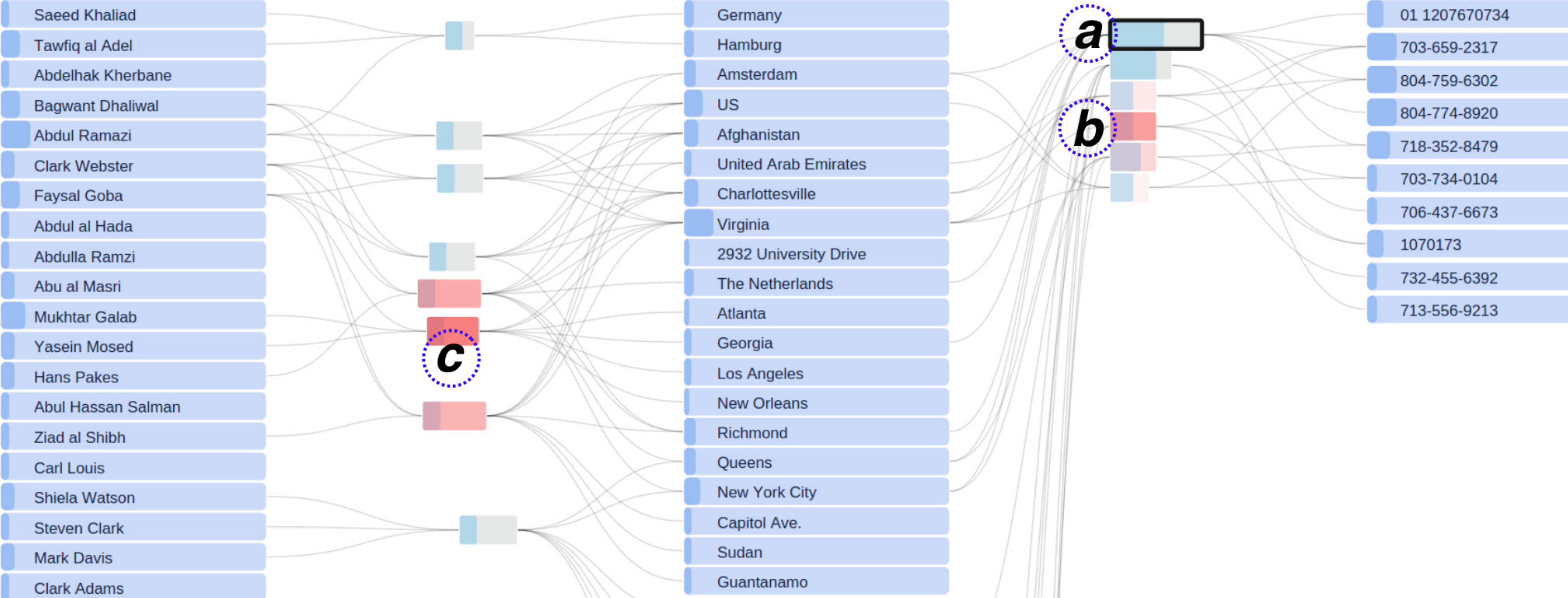}
  \caption{Exampled results from the \textsl{stepwise} evaluation in
  \MaxEntBiSet. (a) shows the bicluster selected by a user to initiate the
maximum
  entropy model evaluation. (b) represents the most surprising bicluster in
  the same bicluster list as the one requested for evaluation. (c) illustrates
  the most surprising bicluster in another bicluster list.}
  \label{fig:biset-step}
\end{figure}

The \textsl{stepwise evaluation} in \MaxEntBiSet examines neighboring biclusters for
the one that users request for evaluation. \textsl{Neighboring biclusters} for a
specific bicluster refers to those that can meet certain degree of overlaps, 
with respect to participated entities, with a user requested bicluster.
\MaxEntBiSet uses the Jaccard coefficient to measure the degree of overlaps
between two biclusters with a default threshold set as $0.1$. 
Thus, for a specific bicluster,
its potential neighboring biclusters are those sharing at least one domain
(e.g., people, location, date, etc.) with this one.

Similar to the \textsl{full path evaluation}, the \textsl{stepwise evaluation}
also has three key steps, including: 1) neighboring bicluster search, 2)
neighboring bicluster evaluation, and 3) neighboring bicluster coloring. Based
on a user specified bicluster for evaluation, \MaxEntBiSet first identifies its
neighboring biclusters using the Jaccard coefficient. Then, \MaxEntBiSet
converts the identified neighboring biclusters into different sets of tiles
following Equation~\eqref{eq:biTiles} and employs the maximum
entropy models to score them. Based on the model evaluation score, BiSet
applies a linear mapping function to assign the opacity value of
\textsl{surprisingness} oriented highlighting color to these biclusters. The
more red a color is, and the higher score this neighboring bicluster gets, which
indicates more surprising information. Figure~\ref{fig:biset-step} gives an
example of the \textsl{stepwise evaluation} in \MaxEntBiSet. In this example,
users request to evaluate a bicluster (see \textsl{a}), \MaxEntBiSet highlights
neighboring biclusters based on their model evaluation scores. Of these
highlighted biclusters, bicluster \textsl{b} shows the most surprising one in
the same bicluster list as that requested for evaluation, and bicluster
\textsl{c} is the most surprising bicluster in the adjacent bicluster list.
Although bicluster \textsl{b} here could not be used to extend the users
selected bicluster \textsl{a}, it has the potential to reveal entities related
to the bicluster \textsl{a} and the plots. Thus, we also take the most
surprising bicluster from the same relation of the users selected bicluster into
account. Such \textsl{stepwise evaluation} potentially enables to involve users
in the process of building a meaningful bicluster-chain. Each time after a
\textsl{stepwise evaluation}, users can investigate highlighted neighboring
biclusters, identify and then select useful one(s) for further exploration.
Users can iterate this process and build a bicluster-chain that is meaningful
for them.

\subsection{Bicluster based Evidence Retrieval}
\begin{figure}[!t]
  \centering
  \includegraphics[width=1\textwidth]{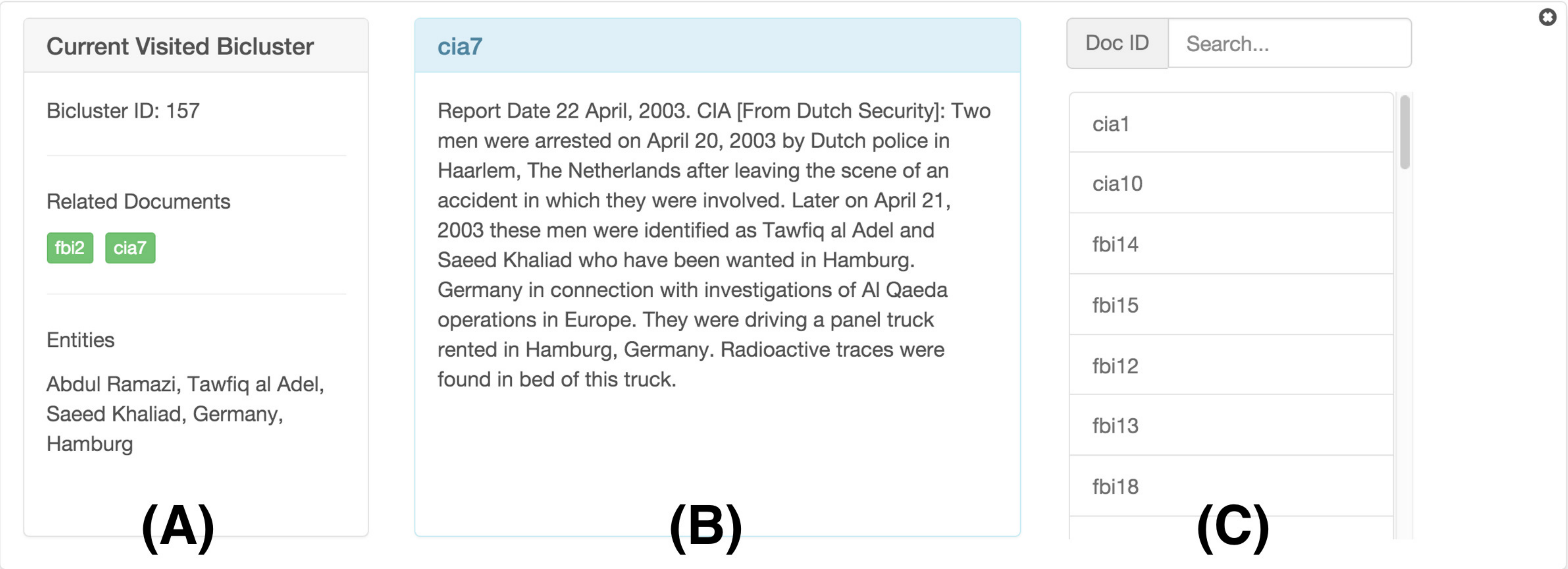}
  \caption{Document view mode in \MaxEntBiSet. (A) depicts the bicluster ID, relevant document
  ID(s) and associated entities. (B) shows the content of a document. (C) lists
  all document IDs in a dataset with a search function.}
  \label{fig:biset-doc}
\end{figure}

\MaxEntBiSet allows users to review relevant documents directly from biclusters
with a right click menu. When investigating a bicluster, users can use a right
click menu to open a popup view where relevant documents are listed, shown in
Figure~\ref{fig:biset-doc}. This helps users review information relevant to
this bicluster and verify computationally identified coalitions of entities. This
document view is on top of the view for relationship exploration with
transparency, so users can simultaneously see both the visualized relationships
and corresponding documents. Moreover, after reading the documents, users can
quickly return to previous view by closing it.

\section{Experiments}
\label{sec:exp}

We describe the experimental results over both synthetic and real datasets. For
real datasets, we focus primarily on datasets from the domain of intelligence
analysis. Through a case study, we demonstrate how the proposed maximum entropy
models embedded in our visual analytics approach
helps analysts to explore text datasets, such as used in intelligence analysis.
All experiments described in this section were conducted on a Xeon 2.4GHz
machine with 1TB memory. Performance results (for synthetic data)
were obtained by averaging over 10 independent runs.

\subsection{Results on Synthetic Data}
\label{sec:exp_syn}

To evaluate the runtime performance of the proposed maximum entropy models with
respect to the data characteristics, we generate synthetic datasets. Since we
focused on the runtime performance of the proposed models here, and the
multi-relational schema of the dataset will not affect how the proposed models
are inferred over the data matrix $D$, we will temporarily ignore the
multi-relational schema of the dataset in the synthetic data for now. The
synthetic datasets are parameterized as follows. The data matrix $D$ consists of
$N$ rows and $M$ columns, or entities, and $\beta$ denotes the density of the
data matrix $D$.  For each entry in the data matrix $D$, we set its value to be
non-zero with probability $\beta$. For the binary case, the non-zero values would
naturally be one, and for the real-valued case, the non-zero values are generated
from a standard uniform distribution. In order to avoid the scenario that too
many rows or columns in $D$ contains only zeros, a non-zero value is placed
randomly in a row or column if it only contains zeros.

\begin{figure}[t]
	\centering
	\includegraphics[width=2.7in]{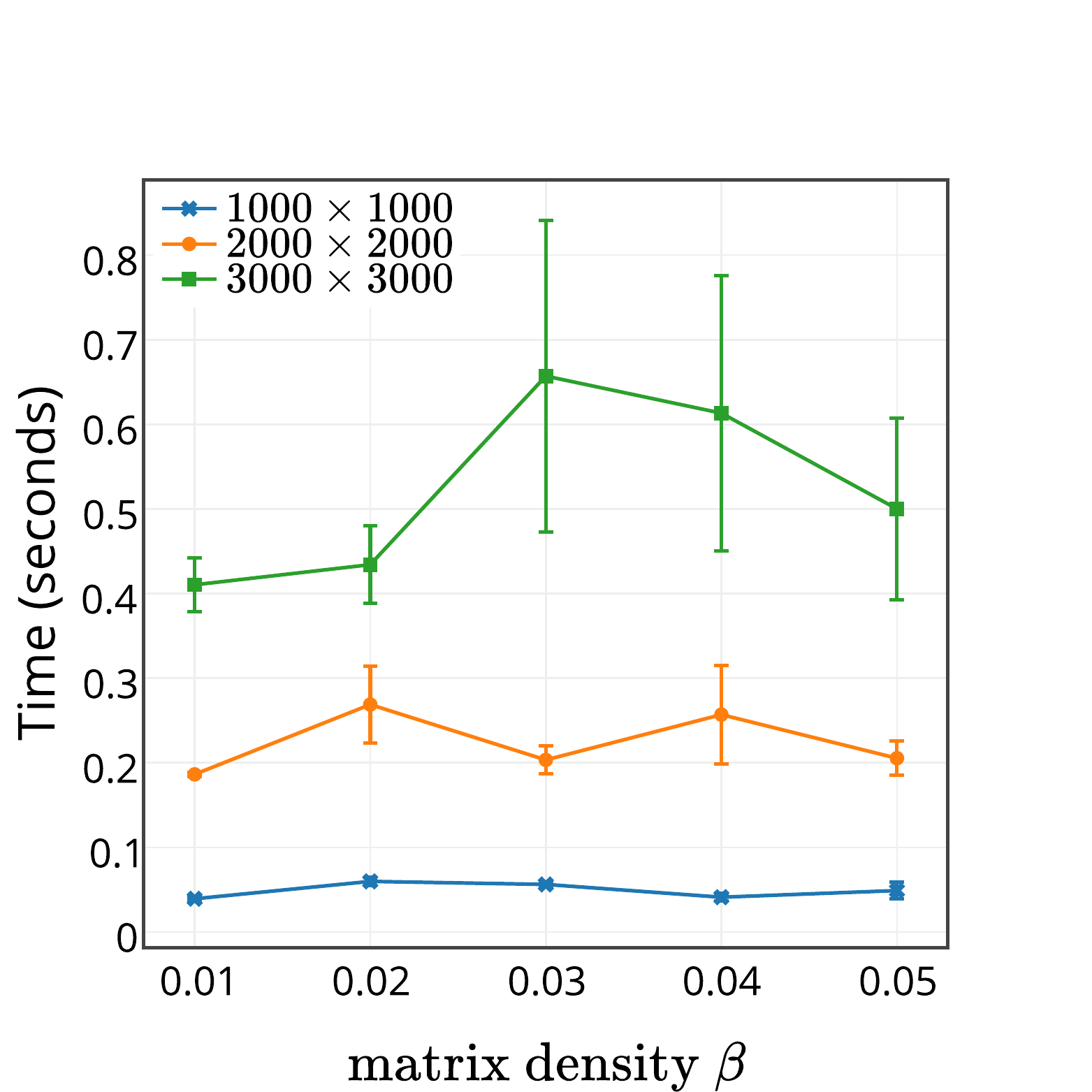}\hfill
	\includegraphics[width=2.7in]{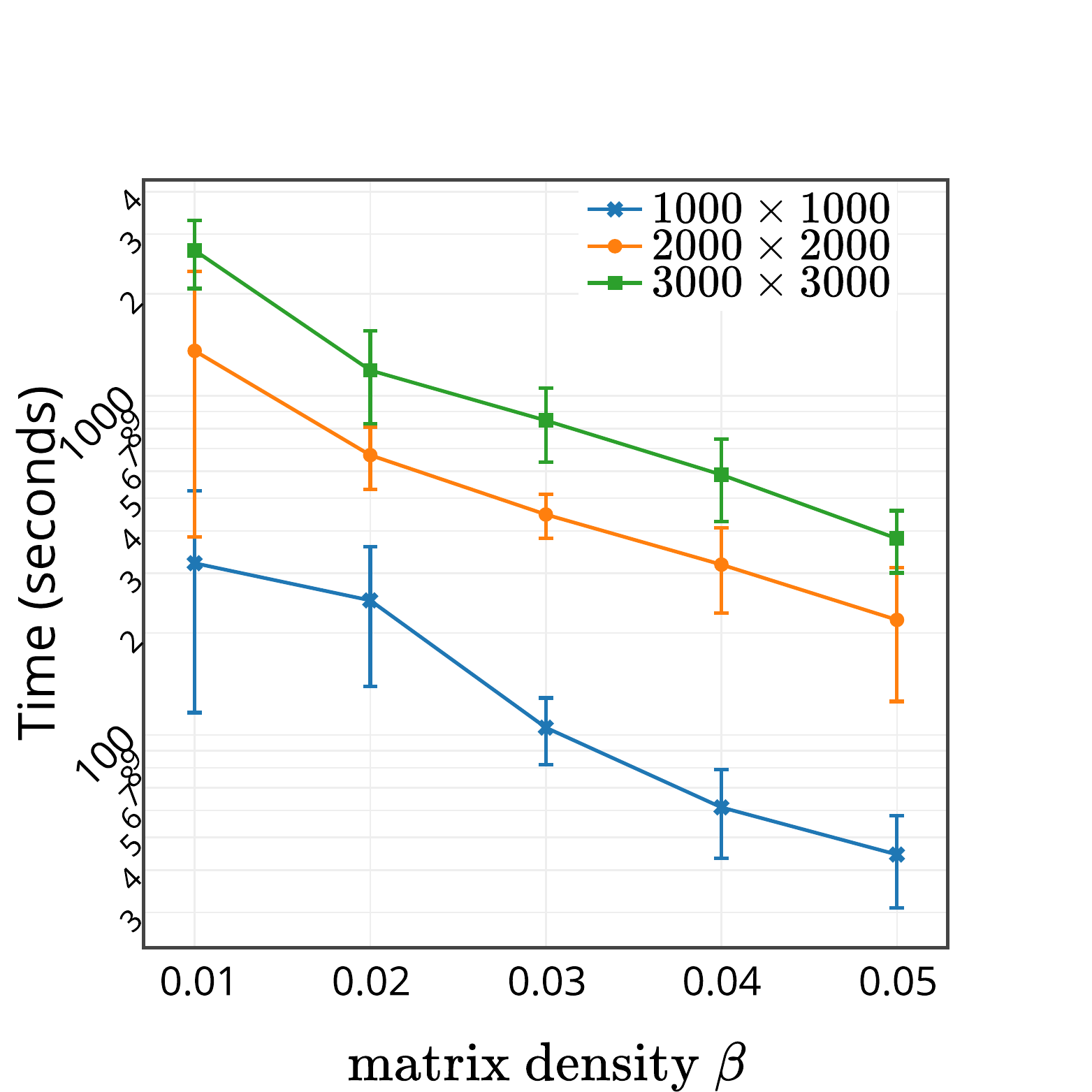}
	\caption{Time to infer the binary (left) and real-valued (right, Y-axis is
	in log scale) maximum entropy model on synthetic
	datasets. The error bars represent the standard deviation}\label{fig:model_infer}
\end{figure}

In our experiments, we explore data matrix $D$ sizes of
($N=1000, M=1000$),
($N=2000, M=2000$), and ($N=3000, M=3000$), and varied the density $\beta$ of the
data matrix $D$ from $0.01$ to $0.05$ in steps of $0.01$. To infer the
maximum entropy models, we use column margin and row margin tiles as the set of
constraint tiles for the proposed model (see Sect.~\ref{sec:model}). We first
investigate the time needed to infer the maximum entropy models.
Figure~\ref{fig:model_infer} shows the model inference time for the binary and
real-valued maximum entropy formulations. As expected, model inference increases with
dataset size and requires more time for the real-valued model.
Since the
real-valued maximum entropy model adopts the conjugate gradient method,
model inference time heavily depends
upon the structure of the given dataset, the number of constraint tiles, and how
fast the model converges to the optimal solution along the gradient direction.
For example, in our experiments we used the row and column margin tiles as the
constraints for the real-valued maximum entropy model, the dimension of the
gradient could be $2 (M + N)$ (that would be 4,000 dimension when $N=1000,
M=1000$ for our synthetic datasets). 

Another interesting phenomenon we observed here is that as the density $\beta$
of the data matrix $D$ increases, the inference time required by the real-valued 
maximum entropy model decreases. One explanation for this phenomenon is that
denser data matrices provide more information to the maximum entropy
model about the underlying data generation distribution through the constraint
tiles. This aids the model in rapidly learning the structure of the data space
and search for the optimal solution with fewer iterations of the conjugate gradient
algorithm.

\begin{figure}[t]
	\centering
	\includegraphics[width=2.7in]{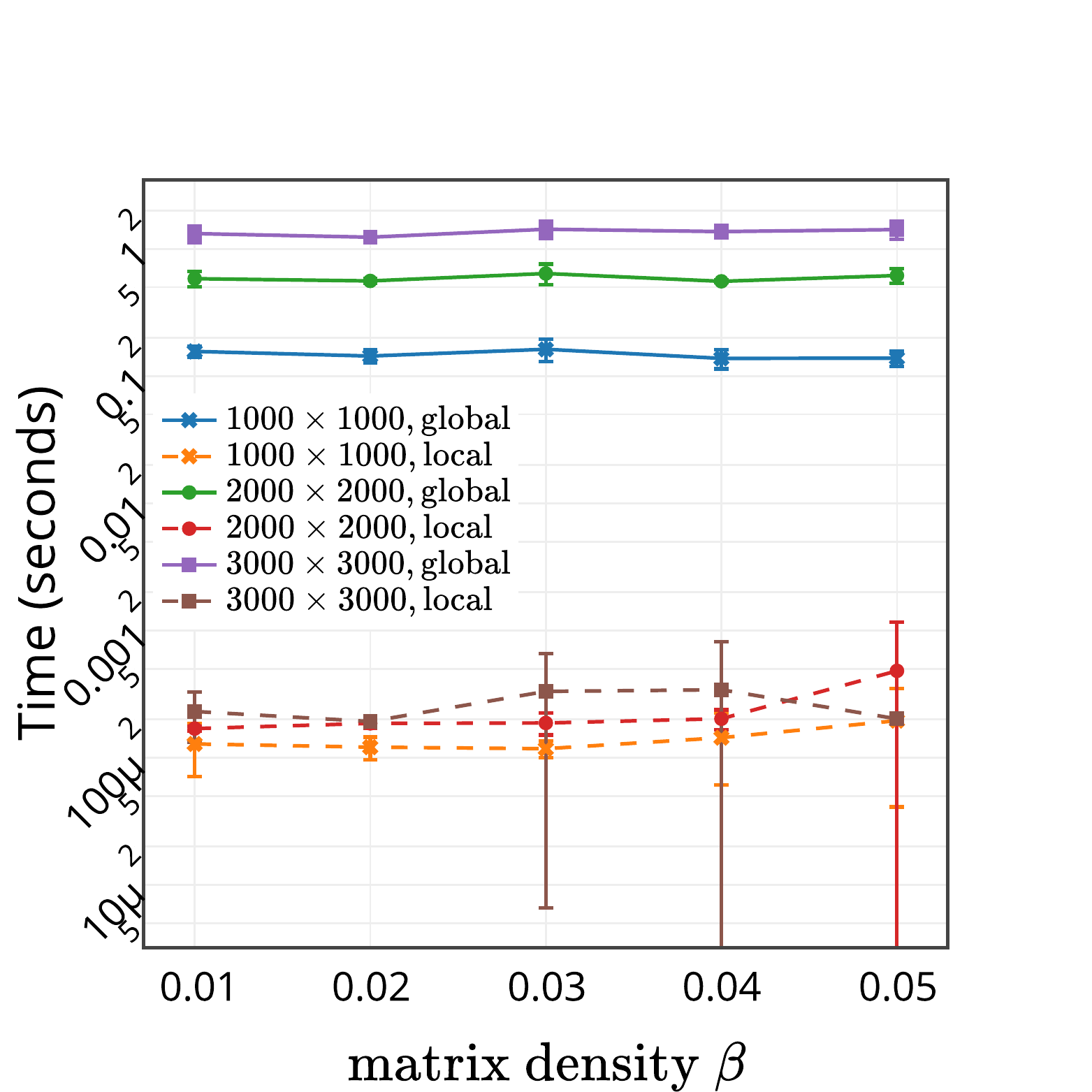}\hfill
	\includegraphics[width=2.7in]{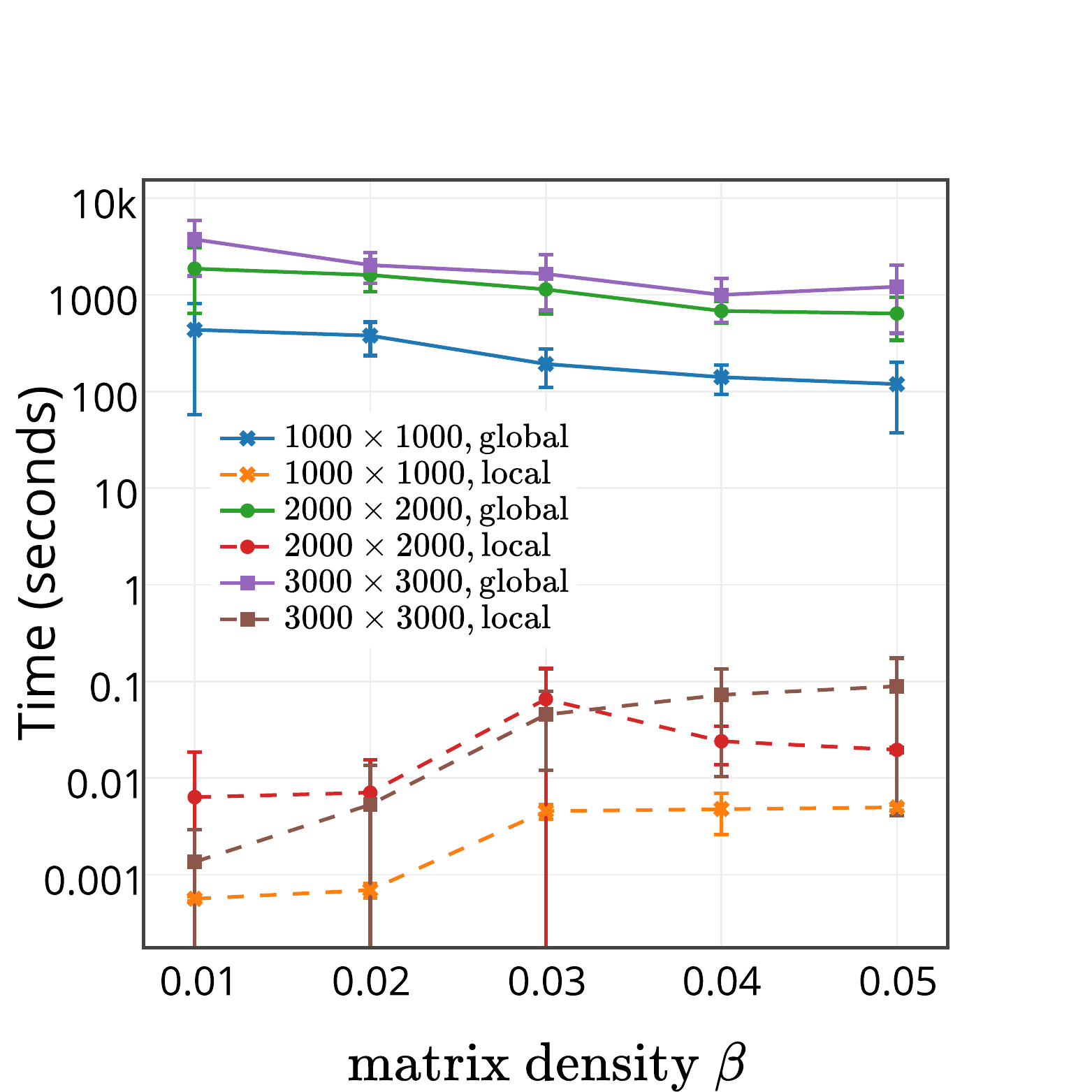}
	\caption{Time to evaluate a set of tiles with the binary (left) and
	real-valued (right) maximum entropy model on synthetic datasets. The set of
	solid lines on the top represents the results of global score, and the set of
	dash lines at the bottom represents the results of local score. The error
	bars represent the standard deviation, and the Y-axis is in log
	scale.}\label{fig:model_eval}
\end{figure}

We also measured the runtime performance of evaluating tile sets with the
proposed binary and real-valued maximum entropy models since the patterns
(biclusters or bicluster chains) whose qualities we would like to assess will
eventually be converted into a set of tiles in our framework. To be more
specific, we randomly generated a set of tiles over the synthetic data matrix,
and compared the time required to evaluate this tile set with both global score
and local score using converged binary and real-valued models, and
Figure~\ref{fig:model_eval} illustrates the results. As we can see from this
figure, in both binary and real-valued maximum entropy model, evaluating tile
sets using the global score requires more time than the local score, which is
expected since the global score requires a complete re-inference of the model.
The difference of runtime performance between global and local scores is significant
in the real-valued model due to this model inference step.
When applying the real-valued maximum entropy model in practical applications,
such as the one here necessitating real-time interaction, we can employ
an asynchronized model inference scheme, e.g.\
creating a daemon process to infer the model when the system is idle, and adopt
the local score to evaluate tile sets. 

\subsection{Evaluation on Real Dataset: A Usage Scenario}
\label{sec:exp_real}

In this section, we walk through an intelligence analysis scenario to
demonstrate how \MaxEntBiSet, particularly incorporating the proposed
maximum entropy models for identifying surprising entity coalitions,
can support an analyst to discover a coordinated activity via visual analysis
of entity coalitions. For ease of description, we use a small
dataset, viz. \textsl{The Sign of the Crescent}
\cite{hughes2003discovery}, which contains $41$ fictional intelligence reports
regarding three coordinated terrorist plots in three US cities where each plot
involves a group of (at least four) suspicious people. In fact, $24$ of the
reports are relevant to the plots. We use LCM~\cite{uno2004efficient} to find
\textsl{closed} biclusters from the dataset with the \textsl{minimum support}
parameter set to 3, which assures that each bicluster has at least three
entities from one domain (e.g., people, location, date, etc.). This leads to 337
biclusters from 284 unique entities and 495 individual relationships (based on
entity co-occurrence in the reports). 

In order to try to discover all the possible plots hidden in the
\textsl{Crescent} dataset, in \MaxEntBiSet, we set the threshold for the Jaccard
coefficient as 0.05, which is a loose constraint. This enables the model to
evaluate those neighboring biclusters that has a few entity overlaps with user
specified biclusters for assessment. Although \MaxEntBiSet fully supports
pattern evaluations with the real-valued \textsl{maximum entropy model}, we
observed that the model evaluation results of a given bicluster were similar
when using the binary and the real-valued \textsl{maximum entropy models} in our
experiments over the \textsl{Crescent} dataset. Thus, we only present the use
case study using the binary \textsl{maximum entropy model} here to demonstrate
the effectiveness of the proposed \MaxEntBiSet technique when assisting analysts
in conducting intelligence analysis tasks.

To illustrate the benefits of integrating the maximum entropy models
into visual analytic tools, in this intelligence analysis scenario, we
use BiSet~\cite{sun2015biset} as the baseline approach for comparison purposes.
Notice that BiSet does not has the capability of model evaluations, and thus it just
provides the \textsl{connection} oriented highlighting function for users to
manually explore entity coalitions. We begin our discussions with the use case
of BiSet, and then discuss the use case of \MaxEntBiSet.

In our scenario, suppose that Sarah is an
intelligence analyst. She is assigned a task to read intelligence reports and
identify potential terrorist threats and key persons from the \textsl{Crescent}
dataset. She opens BiSet, selects four identified domains (people, location,
phone number and date), and begins her analysis.
Figure~\ref{fig:biset-case-without-evaluation} demonstrates Sarah's key
analytical steps using BiSet. Figure~\ref{fig:biset-case-with-evaluation1} and
Figure~\ref{fig:biset-case-with-evaluation2} show the key steps of Sarah's
analytical process using \MaxEntBiSet. 

\subsubsection{BiSet Use Case}

\begin{sidewaysfigure}
  \centering
  \includegraphics[width=0.9\textwidth]{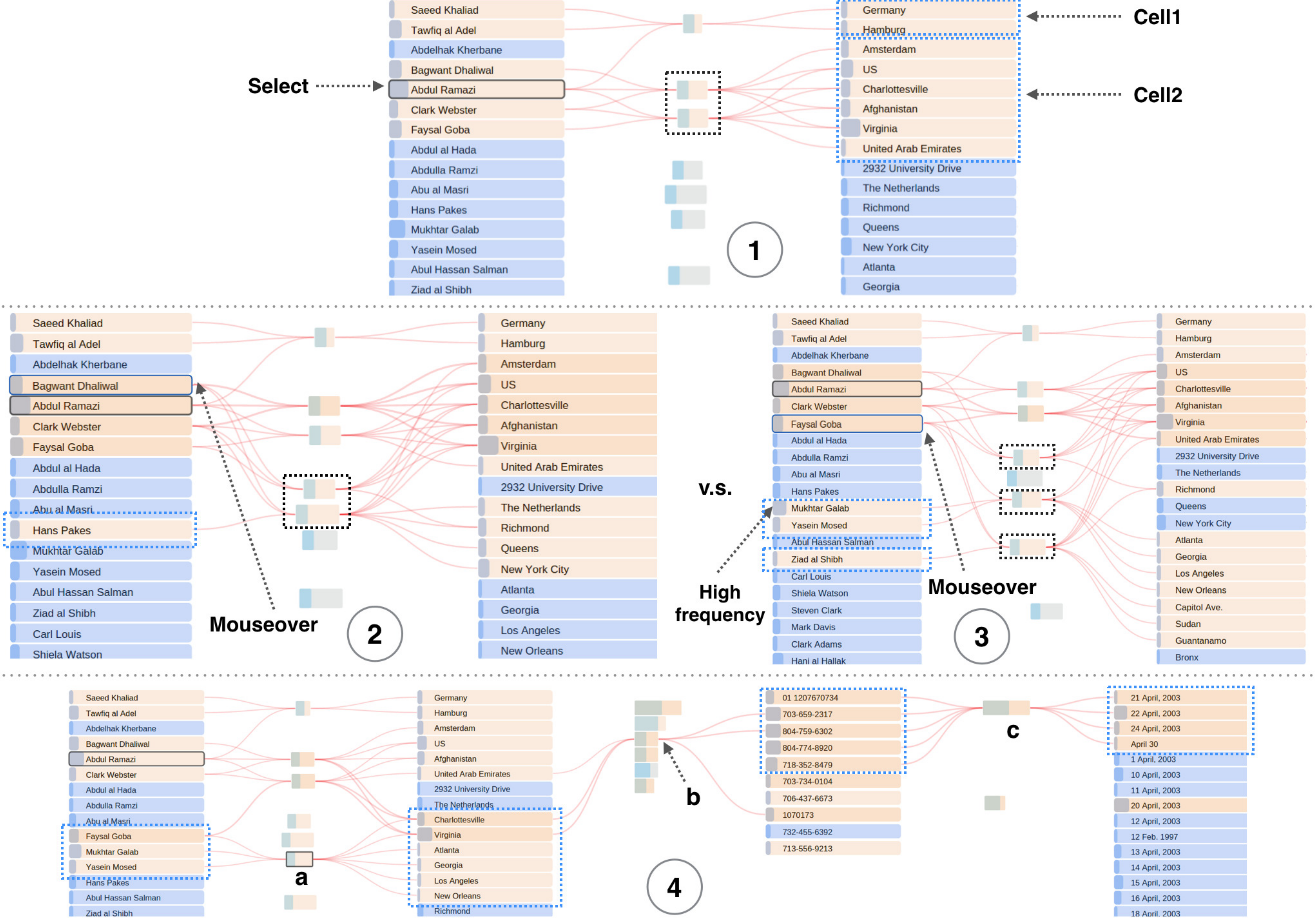}
  \caption{The process of finding a major threat plot with key steps. (1): Based
	  on \textsl{A. Ramazi}, finding that there are two similar bundles and two
	  cells. (2): One name and two bundles are highlighted when hovering the
	  mouse over \textsl{B. Dhaliwal}. (3): Three names and three bundles are
	  highlighted when exploring \textsl{F. Goba}. (4) Referring to the four
  connected groups of useful entities for hypothesis generation.}
  \label{fig:biset-case-without-evaluation}
\end{sidewaysfigure}

Sarah begins analysis by hovering over individual entities in the list of people's
names. BiSet highlights related bundles and entities, each time when she hovers
the mouse over an entity. Immediately she finds that \textsl{A. Ramazi} is
active in three bundles, which indicates that this person may be involved in
three coordinated activities. Sarah selects it
(Figure~\ref{fig:biset-case-without-evaluation} (1)) to focus on highlighted
entities of the three bundles. She finds that \textsl{A. Ramazi} is involved in
two cells with other five people. One is in Germany and the others may be more
broadly located in other four countries.  \textsl{A. Ramazi} is the only person
connecting the two cells, and there are two overlapped subgroups of people
involved in the broader cell.  Moreover, each subgroup has its unique person
(\textsl{B.  Dhaliwal} and \textsl{F. Goba}).

Then Sarah decides to explore the two overlapped subgroups, because she aims to
know what brings the unique people to them. She checks \textsl{B. Dhaliwal}
first by hovering the mouse over it. After this, two bundles are highlighted.
Following edges from them, Sarah finds that they share two people's names and
three locations, but the bigger one (shown in
Figure~\ref{fig:biset-case-without-evaluation} (2)) is related to a new name
(\textsl{H. Pakes}). Then she examines \textsl{F. Goba} in the same way. This
time three bundles and three names are highlighted, and one name (\textsl{M.
Galab}) has a high frequency. This quickly catches Sarah's attention, so she
decides to temporarily pause the analytical branch of \textsl{B. Dhaliwal}, and
moves on with the branch of \textsl{F. Goba}. Sarah hovers the mouse over
\textsl{M. Galab} to check whether it leads to more information. However, it
turns out that no additional bundles or names are highlighted. Sarah realizes
that people potentially related with \textsl{M. Galab} have already been
highlighted in her current view. The bundle (shown in
Figure~\ref{fig:biset-case-without-evaluation} (3) as the black dot box in the
middle) reveals the people related with \textsl{M. Galab}, and all their
activities are in the US\@. With this bundle, Sarah acquires this key insight
revealed by a group of locations. The relations revealed in this bundle are
important, and Sarah infers that the three people (\textsl{M. Galab}, \textsl{Y.
Mosed} and \textsl{Z. al Shibh}) may work on something together in the US\@.
Thus, she decides to find more relevant information by following this tail
\cite{kang2009evaluating}.

Sarah selects the same bundle. BiSet highlights relevant bundles that
potentially form bicluster chains with the selected one. She finds that five
bundles, in the space between the location list and the phone number list, are
highlighted, and two bundles, in the space between the phone number list and the
date list, are highlighted. Relevant entities in lists are also highlighted.

In the two lists of newly highlighted bundles, Sarah finds that there are two
big ones (relatively longer in width shown in
Figure~\ref{fig:biset-case-without-evaluation} (4)) in each list. These two
bundles seem useful since they contain more relations. Sarah chooses to
investigate these first and tries to check how bundles from different
relationship lists are connected. For bundles between the location list and the
phone number list (from top to bottom), Sarah finds that the first bundle and
the third one share two locations (\textsl{Charlottesville} and
\textsl{Virginia}) with the selected bundle, and other highlighted bundles just
share one location with the selected one. Compared with the third bundle, the
first one is related with more locations that are not associated the selected
bundle. Sarah chooses to focus on information highly connected with the selected
bundle, rather than additional information. Thus, she considers the third bundle
a useful one. With the same strategy in another bicluster list, she finds that
the bigger bundle is more useful. 

After this step, Sarah hides edges of other bundles with the right click menu to create
a clear view. Then her workspace shows that three bundles connected to each
other through two shared locations and three shared phone numbers. Sarah feels
that she has found a good number of relations, connecting four groups of
entities, which may reflect a suspicious activity. Therefore, she decides to
read relevant documents to find details of such connections and generate her
hypothesis.

The three connected bundles direct Sarah to eight reports, which are all
relevant to the plot. Sarah reads these reports by referring to the entities
with bright shading in the four connected groups (shown in
Figure~\ref{fig:biset-case-without-evaluation} (4)). The darker shading of an
entity indicates that it is shared more times. Sarah uses this information to help keep her
attention to more important entities in reports. After reading the reports, she
identifies a potential threat with four key persons as follows:

\begin{quote}
\textsl{F. Goba}, \textsl{M. Galab} and \textsl{Y. Mosed}, following the
commands from \textsl{A. Ramazi}, plan to attack \textsl{AMTRAK Train $19$} at
\textsl{9:00 am} on \textsl{April 30}.
\end{quote}

In this use case, Sarah has to manually check details about shared entities to
determine which biclusters are meaningful and useful because BiSet does not
provide the function of model based bicluster-chain evaluation. With just
\textsl{connection} oriented highlighting, Sarah has to verify many connected
biclusters to find potentially useful ones (e.g., finding \textsl{b} in
Figure~\ref{fig:biset-case-without-evaluation} (4) as a useful bicluster). This
limits her analysis strategy as stepwise search, and such search focuses on
checking the shared entities of investigated biclusters. Thus, it takes Sarah
significant effort to work at the
entity-level to finally identify a
meaningful bicluster chain.

\subsubsection{\MaxEntBiSet Use Case}

\begin{sidewaysfigure}
  \centering
  \includegraphics[width=1\textwidth]{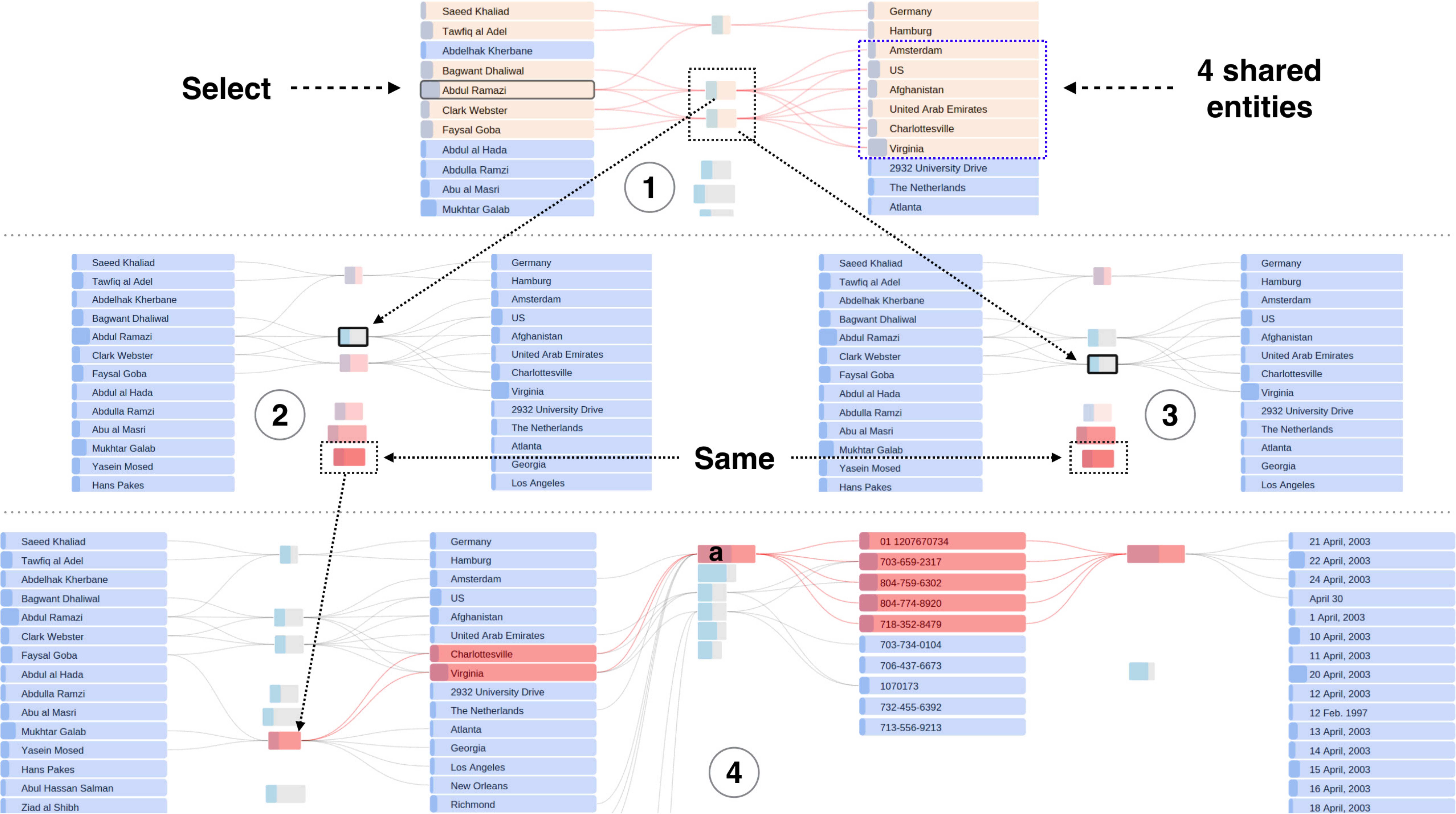}
  \caption{A process of finding one major threat plot with key steps. (1): Based
	  on \textsl{A. Ramazi}, finding that there are two similar bundles. (2):
	  Finding the most surprising bundle evaluated by the \textsl{Maximum
	  Entropy model} for one of the two similar bundles. (3): Finding the most
	  surprising bundle evaluated by the \textsl{maximum entropy model} for the
	  other one of the two similar bundles. (4). The most surprising
	  bicluster-chain suggested by the \textsl{maximum entropy model} for the
   	  most surprising bundle identified in previous steps.}
  \label{fig:biset-case-with-evaluation1}
\end{sidewaysfigure}

Similar to the previous case, Sarah begins analysis by hovering individual
entities in the list of people. \MaxEntBiSet highlights related bundles and
entities as she hovers the mouse over an entity. Immediately she finds that
\textsl{A. Ramazi} is active in three bundles
(Figure~\ref{fig:biset-case-with-evaluation1} (1)), which indicates that this
person is involved in three coordinated activities. Based on edges, Sarah finds
that two bundles are similar (see the black dotted box in
Figure~\ref{fig:biset-case-with-evaluation1} (1)) due to the number of their
shared entities. Thus, she decides to further investigate them.

With the right click menu on the two bundles, Sarah uses the \textsl{stepwise}
evaluation function, provided by \MaxEntBiSet, to find their neighboring bundles
that contain the most surprising information
(Figure~\ref{fig:biset-case-with-evaluation1} (2) and (3)). Based on evaluated
scores from the \textsl{maximum entropy model}, \MaxEntBiSet highlights their
most surprising neighboring bundles. She finds that the most surprising bundles
connected with the two investigated bundles are the same. This indicates that
the model-suggested most surprising bundle may be important and worthy of further
inspection, and so Sarah decides to find more relevant information from it.

Sarah chooses the \textsl{full path} evaluation function on this 
model-suggested bundle to find the most surprising bicluster-chain. \MaxEntBiSet
highlights the path (Figure~\ref{fig:biset-case-with-evaluation1} (4)) passing
through this bundle having the highest evaluation score from the
maximum entropy model. This provides four connected sets of entities
from all the selected domains (people, location, phone and date). Sarah feels
that she has discovered enough information for a story, so she checks entities
involved in this chain and reads documents from the three connected bundles. The
three bundles directs Sarah to nine reports in total, and eight of them are
relevant to each other. After reading these relevant reports, she identifies a
potential threat with four key persons as follows:

\begin{quote}
\textsl{F. Goba}, \textsl{M. Galab} and \textsl{Y. Mosed}, following the
commands from \textsl{A. Ramazi}, plan to attack \textsl{AMTRAK Train $19$} at
\textsl{9:00 am} on \textsl{April 30}.
\end{quote}

Sarah is satisfied with this finding and marks the bundles in this model
suggested chain as useful, using the right click menu. This informs the
integrated maximum entropy model in \MaxEntBiSet that the information
in these bundles has been known to the analyst, and so the model updates its
background information for further evaluations.

\begin{sidewaysfigure}
  \centering
  \includegraphics[width=1.02\textwidth]{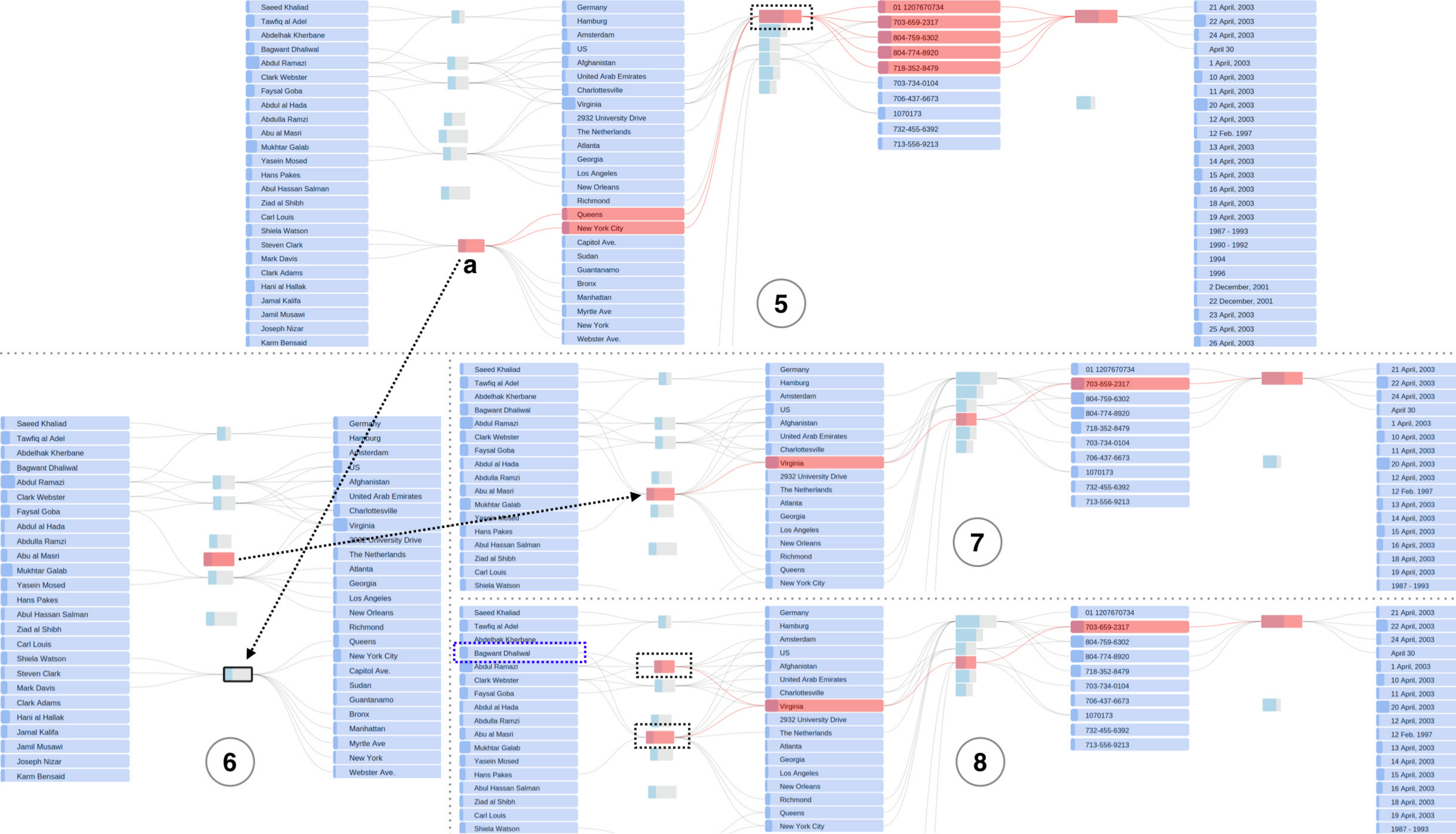}
  \caption{A process of finding another key threat plot. (5): Finding the most
	  surprising bicluster-chain by the \textsl{maximum entropy model} from the
	  bundle marked in the black dotted box. (6): Requesting the
	  \textsl{stepwise} evaluation on the bundle \textsl{a} in (5). (7): Based
	  on the most surprising bundle shown in (6), requesting to find its most
	  surprising bicluster-chain. (8): Based on the shared entity, \textsl{B.
	  Dhaliwal}, requesting to find the most surprising bicluster-chain from the
	  bundle that includes \textsl{B. Dhaliwal} and \textsl{A. Ramazi}. The
  	  chain from this step and that from the previous step merge together.}
  \label{fig:biset-case-with-evaluation2}
\end{sidewaysfigure}

The content of one report, from the bundle in the middle of the surprising chain
(\textsl{a} in Figure~\ref{fig:biset-case-with-evaluation1} (4)), is irrelevant
to that of the other eight, but the entities extracted from this report are
connected with those in the identified threat. Thus, Sarah considers the
information in this report as potentially useful clues, which may lead to some
other threat plot(s). In order to check what new information it can bring in,
she uses the \textsl{full path} evaluation function on the bundle in the
middle of the surprising chain (\textsl{a} in
Figure~\ref{fig:biset-case-with-evaluation1} (4)). Based on this request,
\MaxEntBiSet highlights another chain
(Figure~\ref{fig:biset-case-with-evaluation2} (5)).
This newly highlighted chain has one new bundle (\textsl{a} in
Figure~\ref{fig:biset-case-with-evaluation2} (5)), and this chain merged with
previously suggested surprising chain (comparing
Figure~\ref{fig:biset-case-with-evaluation1} (4) with
Figure~\ref{fig:biset-case-with-evaluation2} (5)). By checking this newly
brought in bundle, Sarah finds that all its entities are different from those in
previously investigated bundles. In order to connect this new piece of
information with previously examined pieces, Sarah decides to use the
\textsl{stepwise} evaluation function on this bundle.

After this stage, \MaxEntBiSet highlights just one bundle
(Figure~\ref{fig:biset-case-with-evaluation2} (6)), which is the most surprising
one suggested by the model. From this bundle, Sarah finds that it includes the
person, \textsl{B. Dhaliwal}. This quickly catches her attention since she
remembers that \textsl{B. Dhaliwal} is connected with \textsl{A. Ramazi}
(Figure~\ref{fig:biset-case-with-evaluation1} (1)). Because of this connection,
Sarah decides to find more information from this bundle and another bundle that
includes \textsl{B. Dhaliwal} and \textsl{A. Ramazi} (the bundle on top in the
black dotted box in Figure~\ref{fig:biset-case-with-evaluation1} (1)), so she
requests the \textsl{full path} evaluation from them. Based on the request from
the newly highlighted bundle shown in
Figure~\ref{fig:biset-case-with-evaluation2} (6), \MaxEntBiSet highlights a new
bicluster-chain (Figure~\ref{fig:biset-case-with-evaluation2} (7)). Then based
on the evaluation request from the bundle including \textsl{B. Dhaliwal} and
\textsl{A. Ramazi}, \MaxEntBiSet highlights another chain. Sarah finds these two
chains merge together (Figure~\ref{fig:biset-case-with-evaluation2} (8)). The
two merged chains both include new pieces of information which connects with the
previous findings. Thus, Sarah decides to read the reports that are related to
these four bundles. 

From the four bundles, in the document view of \MaxEntBiSet, Sarah finds in
total ten unique reports. Of the ten reports, six show evidences about a new
threat and three are those relevant to previously identified threat plot. Based
on the six reports, Sarah identifies the potential threat as:
\begin{quote}
\textsl{B. Dhaliwal} and \textsl{A. Ramazi} plan to attack the \textsl{New York
Stock Exchange} at \textsl{9:00 am} on \textsl{April 30}.
\end{quote}
Considering the connections between this plot and the previously identified one
(e.g., they share some people's names and date), Sarah also confirms that
\textsl{A. Ramazi} is the key person who coordinates the two planned attacks. 

With the capability of model evaluations, in this use case, \MaxEntBiSet
effectively directs Sarah to discover potentially meaningful biclusters or
bicluster-chains. Using colors to visually indicate the model evaluation scores
in \MaxEntBiSet, Sarah can easily see the most surprising bicluster or
bicluster-chain, evaluated by the maximum entropy model. Compared with
the previous use case of BiSet, following the model-suggested biclusters or
chains saves Sarah significant time
in checking entity-level overlaps for meaningful
bicluster identification. In this use case, the maximum entropy model
shares the burden of Sarah for foraging information (e.g., finding potentially
useful biclusters or chains). Thus, compared with the first use case, Sarah can
spend more time and effort to synthesize the visualized structured information
for hypothesis generation.

\subsubsection{Comparison between BiSet and \MaxEntBiSet} 
Both BiSet and \MaxEntBiSet can highlight entities and biclusters based on
connections, and visually present entities and biclusters (algorithmically
identified structured information) in an organized manner. However, compared
with BiSet, \MaxEntBiSet also enables the
highlighting entities and biclusters based
on identified surprising coalitions from the maximum entropy model.
Comparing the two cases discussed above, we find that \MaxEntBiSet better
supports the user's sensemaking process of exploring entity coalitions, than BiSet
does, from two key aspects: 1) efficiency and 2) exploring new analytical paths.

Compared with BiSet, \MaxEntBiSet more effectively directs users' attention to
potentially useful biclusters or bicluster-chains by visually prioritizing them
with colors based on their maximum entropy model evaluation scores. The model
evaluation function provided in \MaxEntBiSet eases the process for users to find
useful biclusters, particularly compared with manually entity overlap
investigation. For example, in the first use case, a user has to examine in
total 9 biclusters ($4$ in the left most bicluster list, $5$ in the middle
bicluster list and $2$ in the right most bicluster list as shown in
Figure~\ref{fig:biset-case-without-evaluation}), before she finally identifies a
meaningful bicluster-chain that covers the information of a potential threat.
However, in the second use case, \MaxEntBiSet directs the user to a
bicluster-chain after she investigates 3 biclusters (in the left most bicluster
list shown in Figure~\ref{fig:biset-case-with-evaluation1}). Although this chain
is slightly different from the manually identified one in the first use case, it
covers the same amount of information as the other one does. Thus, in the second
use case, \MaxEntBiSet saves the user from checking highlighted biclusters in
the other two lists, and effectively provides a useful bicluster-chain for users
to explore.

Based on the four user selected domains (visualized as a fixed schema), it is
hard to identify all three threat plots in the \textsl{Crescent} dataset because
not all pre-identified biclusters can be shown. However, from the two cases, we
can find that \MaxEntBiSet can direct users from one identified plot to a new
plot via a surprising bicluster-chain. However, when users manually forage
relevant information, it is not easy for them to make such transitions due to
cognitive tunneling \cite{thomas2001visual}. In the first use case, the key
bundle that can lead to a new plot is actually identified not as useful as the
one shown as \textsl{b} in Figure~\ref{fig:biset-case-without-evaluation}. Thus,
\MaxEntBiSet significantly aids in identifying coalitions of entities worthy of
further exploration.

\section{Related Work}
\label{sec:related}
In this section we survey related work. In particular, we discuss work related
with regard to mining surprising patterns, iterative data mining,
mining multi-relational datasets, finding plots in data, and bicluster
visualizations for data exploration.



\subsection{Mining Biclusters}
Mining biclusters is an extensively studied area of data mining, and many
algorithms for mining biclusters from varied data types have been
proposed, e.g.~\citep{Tibshirani99clusteringmethods, Califano00analysisof,
Segal01062001, Sheng27092003, Cheng:2000:BED:645635.660833, Zaki:1401887,
Uno:2005:LVC:1133905.1133916}. Bicluster mining, however, is not the primary aim
in this paper; instead it is only a component in our proposed framework.
Moreover, the above mentioned studies do not assess whether the mined clusters
are subjectively interesting. A comprehensive survey of biclustering algorithms
was given by \citet{Madeira:2004:BAB:1024308.1024313}.

\subsection{Mining Surprising Patterns}
There is, however, significant literature on mining
representative/succinct/sur-prising patterns
\citep[e.g.,][]{Kiernan:2009:CCS:1631162.1631169} as well as on explicit
summarization \citep[e.g.,][]{conf:sdm:DavisST09}. \citet{wang:06:summaxent}
summarized a collection of frequent patterns by means of a row-based MaxEnt
model, heuristically mining and adding the most significant itemsets in a
level-wise fashion. \citet{tatti:06:computational} showed that querying such a
model is PP-hard. \citet{mampaey:12:mtv} gave a convex heuristic, allowing more
efficient search for the most informative set of patterns.
\citet{debie:11:dami} formalized how to model a binary matrix by MaxEnt using
row and column margins as background knowledge, which allows efficient
calculation of probabilities per cell in the matrix. \citet{real:value:maxent}
first proposed a real-valued MaxEnt model for assessing patterns over
real-valued rectangular databases. These papers all focus on mining surprising
patterns from a single relation. They do not explore the multi-relational
scenario, and can hence not find connections among surprising patterns from
different relations---the problem we focus on. 

\subsection{Iterative Data Mining} 
Iterative data mining as we study was first proposed by
\citet{hanhijarvi:09:tell}. The general idea is to iteratively mine the result
that is most significant given our accumulated knowledge about the data. To
assess significance, they build upon the swap-randomization approach of
\citet{gionis:07:assessing} and evaluate empirical p-values. With the help of
real-valued MaxEnt model,~\citet{konto:13:numaxentit} proposed a subjective
interestingness measure called \textit{Information Ratio} to iteratively
identify and rank the interesting structures in real-valued data.
\citet{mampaey:12:mtv} and \citet{konto:13:numaxentit} show that ranking results
using a static MaxEnt model leads to redundancy in the top-ranked results, and
that iterative updating provides a principled approach for avoiding this type of
redundancy.  \citet{tatti:12:apples} discussed comparing the informativeness of
results by different methods on the same data. They gave a proof-of-concept for
single binary relations, for which results naturally translate into tiles, and
gave a MaxEnt model in which tiles can be incorporated as background knowledge.
In this work we build upon this framework, translating bicluster chains (over
multiple relations) into tiles to measure surprisingness with regard to
background knowledge using a Maximum Entropy model.

\subsection{Multi-relational Mining} 
Mining relational data is a rich research area~\citep{rdmbook} with a plethora
of approaches ranging from relational association
rules~\citep{Dehaspe:2001:DRA:567222.567232} to inductive logic programming
(ILP) \citep{ilp}.  The idea of composing
redescriptions~\citep{Zaki:2005:RSU:1081870.1081912} and biclusters to form
patterns in multi-relational data was first proposed by
\citet{Jin:2008:CMM:1342320.1342322}.  \citet{Cerf:2009:CPM:1497577.1497580}
introduced the \textsc{DataPeeler} algorithm to tackle the challenge of directly
discovering closed patterns from $n$-ary relations in multi-relational data.
Later, \citet{Cerf:2013:noise:tolerant} refined \textsc{DataPeeler} for finding
both closed and noise-tolerant patterns.  These frameworks do not provide any
criterion for measuring subjective interestingness of the multi-relational
patterns.  \citet{DBLP:conf/sdm/OjalaGGM10} studied randomization techniques for
multi-relational databases with the goal to evaluate the statistical
significance of database queries. \citet{spyropoulou:11:multirel} and
\citet{DBLP:SpyropoulouBB14} proposed to transform a multi-relational database
into a $K$-partite graph, and to mine maximal complete connected subset (MCCS)
patterns that are surprising with regard to a MaxEnt model based on the margins
of this data. \citet{Spyropoulou:local:patterns} extended this approach to
finding interesting local patterns in multi-relational data with $n$-ary
relationships. Bicluster chains and MCCS patterns both identify redescriptions
between relations, but whereas MCCS patterns by definition only identify exact
pair-wise redescriptions (completely connected subsets), bicluster chains also
allow for approximate redescriptions (incompletely connected subsets). All
except for the most simple bicluster chains our methods discovered in the
experiments of Section~\ref{sec:exp} include inexact redescriptions, and could
hence not be found under the MCCS paradigm. Besides that we consider two
different data models, another key difference is that we iteratively update our
MaxEnt model to include all patterns we mined so far.  

\subsection{`Finding Plots'} 
The key difference between finding plots, and finding biclusters or surprising
patterns is the notion of chaining patterns into a chain, or plot.  Commercial
software such as {\it Palantir} provide significant graphic and visualization
capabilities to explore networks of connections but do not otherwise automate
the process of uncovering plots from document collections.
\citet{shahaf-guestrin-journal} studied the problem of summarizing a large
collection of news articles by finding a chain that represents the main events;
given either a start or end-point article, their goal is to find a chain of
intermediate articles that is maximally coherent. In contrast, in our setup we
know neither the start nor end points. Further, in intelligence analysis, it is
well known that plots are often loosely organized with no common all-connecting
thread, so coherence cannot be used as a driving criterion. Most importantly, we
consider data matrices where a row (or, document) may be so sparse or small
(e.g., 1-paragraph snippets) that it is difficult to calculate statistically
meaningful scores. Storytelling algorithms
\citep[e.g.,][]{Hossain:2012:SEN:2339530.2339742,storytelling-tkde,connectingpubmed}
are another related thread of research; they provide algorithmic ways to rank
connections between entities but do not focus on entity coalitions and how such
coalitions are maintained through multiple sources of evidence.
\citet{Wu:2014:UPD:2664051.2664089} proposed a framework to discover the plots
by detecting non-obvious coalitions of entities from multi-relational datasets
with Maximum Entropy principle and further support iterative, human-in-the-loop,
knowledge discovery. However, no visualization framework was developed to enable
analysts to be involved when discovering the surprising entity coaliations in
that work. Moreover, we also propose the full path and step-wise chain search
strategies and combine them together to help analyts to explore the data.

\subsection{Bicluster Visualizations}
Finally, we give an overview of work on bicluster visualization techniques.
Biclusters offer a usable and effective way to present coalitions among sets of
entities across multiple domains. Various visualizations have been proposed to
present biclusters for sensemaking of data in different fields. One typical
application domain of bicluster visualizations is bioinformatics, where
biclusters are visualized to help bioinformaticians to identify groups of genes
that have similar behavior under certain groups of conditions (e.g., BicAt
\cite{barkow2006bicat}, Bicluster viewer \cite{heinrich2011bicluster},
BicOverlapper 2.0 \cite{santamaria2014bicoverlapper}, BiGGEsTS
\cite{gonccalves2009biggests}, BiVoc \cite{grothaus2006automatic}, Expression
Profiler \cite{kapushesky2004expression}, GAP \cite{wu2010gap} and Furby
\cite{streit2014furby}). In addition, \citet{fiaux2013bixplorer} and
\citet{sun2014role} applied biclusters in Bixplorer, a visual analytics tool, to
support intelligence analysts for text analytics. Evaluations of these tools
show promising results, which indicates that using visualized bicluster to empower
data exploration is beneficial.

In order to systematically inform the design of bicluster visualizations, a
five-level design framework has been proposed \cite{sun2014five} and the key
design trade-off to visualize biclusters has been identified:
\textsl{Entity-centric} and \textsl{relationship-centric} \cite{sun2015biset}.
This design framework highlights five levels of relationships that underlie the
notions of biclusters and bicluster chains. The design trade-off suggests that
bicluster visualizations should visually represent both the membership of
entities and the overlap among biclusters in a human perceptible and usable
manner.

\section{Conclusion}
\label{sec:conclusion}
Our approach to discover multi-relational patterns with maximum entropy models
in a visual analytics tool is a significant step in formalizing a previously
unarticulated knowledge discovery problem and supporting its solution in an
interactive manner.  We have primarily showcased results in intelligence
analysis; however, the theory and methods presented are applicable for analysis
of unstructured or discrete multi-relational data in general---such as for
biological knowledge discovery from text. The key requirement to apply our
methods is that the data should be transformed into our data model. 

Some of the directions for future work include (i) obviating the need to mine
all biclusters prior to composition, (ii) improving the scalability of the
proposed models and framework to be able to deal with even larrger datasets,
(iii) enabling dynamic and flexible multi-relational schema generation to
support better sensemaking and hidden plot discovery, (iv) incorporating weights
on relationships to account for differing veracities and trustworthiness of
evidence. Ultimately, the key is to support more expressive forms of
human-in-the-loop knowledge discovery.

\bibliographystyle{plainnat}
\bibliography{bib-jilles,reference,vis-bic}

\end{document}